\documentclass[letterpaper,twocolumn,10pt]{article}
\usepackage{usenix}

\usepackage{graphicx}

\usepackage{tikz}
\usepackage{amsmath}

\usepackage{booktabs}
\usepackage{subfigure}
\usepackage{url} 
\usepackage{comment}
\usepackage{accsupp}
\usepackage{hyperref}
\usepackage{breakurl}  

\providecommand{\eg}{\emph{e.g.,} }
\newcommand\mypara[1]{\noindent \textbf{#1}}

\begin{document}

\date{}

\title{\Large \bf A Midsummer Meme’s Dream:\\ Investigating Market Manipulations in the Meme Coin Ecosystem}

\author{
{\rm Alberto Maria Mongardini}\\
Sapienza University of Rome, Technical University of Denmark\\
mongardini@di.uniroma1.it, among@dtu.dk\\
\and
{\rm Alessandro Mei}\\
Sapienza University of Rome\\
mei@di.uniroma1.it\\
} 

\maketitle

\begin{abstract}
From viral jokes to a billion-dollar phenomenon, meme coins have become one of the most popular segments in cryptocurrency markets. Unlike utility-focused crypto assets like Bitcoin, meme coins derive value primarily from community sentiment, making them vulnerable to manipulation.
This study presents an unprecedented cross-chain analysis of the meme coin ecosystem, examining 34,988 tokens across Ethereum, BNB Smart Chain, Solana, and Base. We characterize their tokenomics and track their growth in a three-month longitudinal analysis. 
We discover that among high-return tokens (>100\%), an alarming 82.8\% show evidence of artificial growth strategies designed to create a misleading appearance of market interest. These include wash trading and a new form of manipulation we define as Liquidity Pool-Based Price Inflation (LPI), where small strategic purchases trigger dramatic price increases.
We find that profit extraction schemes, such as pump and dumps and rug pulls, typically follow initial manipulations like wash trading or LPI, indicating how early manipulations create the foundation for later exploitation.
We quantify the economic impact of these schemes, identifying over 17,000 victimized addresses with realized losses exceeding \$9.3 million.
These findings reveal that combined manipulations are widespread among high-performing meme coins, suggesting that their dramatic gains are often driven by coordinated efforts rather than natural market dynamics. 
\end{abstract}

\section{Introduction}
The explosive rise of meme coins represents one of the most impressive trends in cryptocurrency markets. What started as a joke with Dogecoin in 2013~\cite{nani2022doge} has evolved into a market worth over \$117 billion by 2024~\cite{meme_success_cmc}. 
Their appeal lies in the potential for extreme returns, exemplified by Pepe (400,000\%) and Shiba Inu (80,000,000\%)~\cite{impressive_return_meme_coins_spa}. These extraordinary profits have attracted waves of investors hoping to find the next meme coin before its price explodes, leading to an unprecedented proliferation of meme coins, largely driven by platforms like pump.fun, which allow users to create tokens with minimal technical skills~\cite{pump_success}.
The success of meme coins has prompted regulatory institutions to develop specialized definitions, underscoring their importance as a distinct asset. The U.S. Securities and Exchange Commission (SEC) defines meme coins as "a type of crypto asset inspired by internet memes, characters, current events, or trends for which the promoter seeks to attract an enthusiastic online community to purchase the meme coin and engage in its trading"~\cite{meme_coins_sec}, a definition we adopt in this study.
However, meme coins present a concerning vulnerability: Unlike utility tokens, meme coins rely exclusively on perceived interest to maintain their value~\cite{nani2022doge}, creating strong incentives for manipulations.

This study investigates whether token creators and early investors employ artificial growth strategies to create illusory activity to attract investors, and how these strategies facilitate subsequent profit extraction schemes like pump and dumps~\cite{xu2019anatomy} or rug pulls~\cite{zhou2024stop}. 
First, we collect and publicly release~\cite{meme_coin_repository} a dataset of 34,988 meme coins across four major blockchains (Ethereum, BNB Smart Chain (BSC), Solana, and Base), which, to our knowledge, is the largest curated dataset of meme coins to date.
Our investigation begins with the characterization of meme coin tokenomics across different blockchains.
We then perform a three-month longitudinal analysis of price growth, identifying 707 high-return tokens (>100\%). We uncover an alarming 82.8\% of these high-performing meme coins exhibit evidence of market manipulation, including recurrent wash trading and what we define as Liquidity Pool-Based Price Inflation (LPI), an operation where strategic purchases with minimal capital trigger dramatic price increases. Furthermore, we find profit extraction schemes, such as pump and dumps and rug pulls, designed to capitalize on artificially generated interest: 62.9\% of tokens subjected to these operations had previously undergone wash trading or LPI, revealing how initial manipulations often serve as preliminary stages in broader exploitation strategies.
This connection between manipulations is even stronger among meme coins delisted by established aggregators: 86.67\% of the delisted tokens subjected to profit extraction had previously been artificially inflated. In such cases, token creators first use artificial growth tactics to meet listing requirements, then leverage their newfound visibility to execute exit strategies, maximizing their profits. 
We link these schemes to quantifiable financial harm, documenting \$3.27 million in realized losses from pump and dump schemes and \$6.04 million from rug pulls, affecting over 17,000 victim addresses.
These findings highlight the prevalence of manipulation strategies among high-return meme coins, underscoring the need for enhanced investor protection mechanisms. 

Our main contributions are the following:

\begin{itemize}
    \item \textbf{Curated Dataset of Meme Coins}: We build and publicly release~\cite{meme_coin_repository} the first large-scale dataset of 34,988 meme coins across Ethereum, BNB Smart Chain, Solana, and Base, providing a valuable resource for future works on the meme coins ecosystem.
    \item \textbf{Cross-chain Characterization}: Leveraging this dataset, we conduct the first cross-chain analysis of meme coins, uncovering blockchain-specific patterns in tokenomics, temporal evolution, and price growth.
    \item \textbf{Artificial Growth Detection}: We identify strategies that create misleading market interest, including a new manipulation we define as Liquidity Pool-Based Price Inflation (LPI). We show that 82.8\% of high-performing tokens exhibit manipulation indicators (\eg concentrated ownership, wash trading, or LPI), suggesting that their growth is often due to coordinated manipulation.
    \item \textbf{Profit Extraction and Economic Impact}: We show that profit extraction is often sequential, with 62.9\% of high-return tokens (86.67\% of delisted ones) using artificial growth to facilitate exit scams.
    We quantify the direct cost of these schemes, documenting over \$9.3 million in losses impacting over 17,000 addresses.

    
    
\end{itemize}

\section{Related Work}
\mypara{Memes and Meme Coins.}
Internet memes, defined by Shifman~\cite{shifman2013memes} as ``units of popular culture that are circulated, imitated, and transformed by individual Internet users,'' have become significant cultural phenomena. 
The intersection of meme culture and cryptocurrencies created meme coins, which differ from traditional cryptocurrencies through their community-driven nature, humor incorporation, and meme culture origins~\cite{stencel2023meme, nani2022doge}. 
Kim et al.~\cite{kim2024identifying} studied how social media drives meme coin valuation, while Belcastro et al.~\cite{belcastro2023enhancing} highlighted that meme coin price movements require specialized prediction approaches, different from other crypto assets.
Long et al. introduced CoinCLIP~\cite{long2024coinclip}, a framework leveraging visual-language models to classify meme coin viability, and released the Coin-Meme dataset~\cite{long2024bridging} containing multimodal data from Solana meme coins. 

\mypara{Market Manipulations on Blockchains.}
The cryptocurrency ecosystem, characterized by its relative lack of regulatory oversight, has become susceptible to various forms of market manipulation~\cite{gandal2018price, krafft2018experimental, cernera2023ready, daian2020flash, cernera2024blockchain}. These practices undermine market integrity and pose significant risks to investors.

Wash Trading is a manipulation that aims to create an illusion of market activity by generating artificial volume. Victor and Weintraud~\cite{victor2021detecting} found that up to 35\% of ERC-20 token volume on some exchanges was generated through wash trading, while Le Pennec et al.~\cite{le2021wash} proposed machine learning approaches to identify wash trading activities involving ERC-20 tokens. This manipulation has also extended to non-fungible token (NFT) markets, with~\cite{von2022nft, la2023game} documenting significant wash trading activity in NFT marketplaces.

Pump and dumps represent one of the most prevalent manipulations in cryptocurrency markets. These schemes involve inflating an asset's price through misleading statements and coordinated buying (the "pump"), followed by selling the overvalued assets (the "dump") once enough investors have been lured in~\cite{kamps2018moon}. Xu and Livshits~\cite{xu2019anatomy} developed detection mechanisms for pump and dumps, 
while other studies~\cite{hamrick2019economics,la2023doge,la2020pump} found that pump announcements typically generated average returns of 65\% for organizers, mostly at the expense of unsuspecting investors. These works also examined the communication platforms used to coordinate such schemes, identifying Telegram and Discord as the main hubs, a finding also supported by Telegram-focused research~\cite{la2023sa,la2018pretending,la2025tgdataset}.

Lastly, rug pulls have become a harmful exit scam in decentralized finance (DeFi), where developers abandon projects after attracting investors. Zhou et al.~\cite{zhou2024stop} analyzed 201 rug pulls and developed tools to detect associated risks, while Cernera et al.~\cite{cernera2023ready} identified "1-day rug pulls" that generated \$240 million in profits, primarily on BNB Smart Chain.

This work addresses several critical gaps in existing literature. First, we focus on meme coins, which possess unique characteristics that distinguish them from other cryptocurrency assets. These differences have led regulatory institutions like the SEC to develop specialized definitions for meme coins~\cite{meme_coins_sec}, highlighting the need for dedicated analysis. Second, while previous studies examine individual manipulation types such as wash trading~\cite{la2023game,le2021wash} or pump-and-dumps~\cite{xu2019anatomy,la2020pump}, we analyze how these different strategies are combined and executed sequentially to maximize profits. Finally, we provide a comprehensive cross-chain analysis of the meme coin ecosystem across four major blockchains (Ethereum, BSC, Solana, and Base), surpassing the scope of previous works, which examine at most two blockchains, and offering a more complete overview of market manipulations.

\section{Methodology}
To capture a comprehensive view of the meme coin ecosystem, we design a methodology comprising three steps. We first identify tokens labeled as meme coins by established platforms, then expand our dataset by collecting newly created tokens from additional sources, and finally apply classification techniques to distinguish meme coins from other assets.

\subsection{Data Collection}
The meme coin ecosystem spans multiple blockchains, including well-known tokens and emerging projects. To capture this diversity, we adopt a blockchain-agnostic data collection strategy, sourcing information from mainstream crypto tracking platforms and specialized decentralized exchange (DEX) aggregators.
This cross-chain approach enables us to track migration patterns and observe how meme coin creators leverage different blockchains, uncovering cross-chain trends and insights that would be missed in a single-chain analysis.

\subsubsection{Known Meme Coins}
\label{subsubsec:know_meme_coins}
Our first approach focuses on established cryptocurrency aggregators that explicitly categorize tokens as meme coins, providing a foundation for verified meme coins. For this purpose, we collect data from three primary sources.

We start with CoinMarketCap~\cite{coinmakretcap} and CoinGecko~\cite{coingecko}, two leading cryptocurrency tracking platforms that maintain specialized meme token categories. CoinGecko defines meme coins as ``Tokens with no intrinsic value, but with a strong community that fosters social interaction''~\cite{meme_coins_definition_coingecko}, while CoinMarketCap describes them as ``a type of cryptocurrency that originated from internet memes. They are often created as a joke or for the purpose of satire... Some of these coins have gained popularity due to their community-driven nature and the humorous themes they embody''~\cite{meme_coins_definition_coinmarketcap}. Both definitions align with the one provided by the SEC~\cite{meme_coins_sec}, which we adopt as the formal definition for our work. Using their free APIs~\cite{coinmakretcap_api,coingecko_api}, we collect data on 2,930 and 2,107 meme coins, respectively.
Additionally, we utilize Gecko Terminal~\cite{geckoterminal}, which categorizes tokens by meme trends. This approach reflects the definition of meme coins as assets inspired by Internet trends and culture~\cite{meme_coins_sec}. We develop a custom scraper to collect data on 4,735 meme coins from pools tagged with relevant meme categories, including animal, dog, cat, inu, and pump.fun (a platform for meme coin creation).
By aggregating the results of these services and removing duplicates by identifying each token using the tuple $(token\_contract\_address, blockchain)$, we obtain a total of 8,852 distinct meme coins.

\subsubsection{Searching other Meme Coins}
\label{subsubsec:new_dex_meme_coins}
To expand our dataset, we implement a second approach focusing on newly created tokens. We develop scrapers for two DEX aggregators: DexScreener~\cite{dexscreener} and CoinSniper~\cite{coinsniper}, which track real-time data on emerging tokens across decentralized exchanges. These platforms differ from CoinMarketCap and CoinGecko in their listing criteria. While the latter require stricter conditions, such as functional websites and listing on integrated exchanges~\cite{coingecko_listing, coinmarketcap_listing}, DEX aggregators implement more permissive policies: DexScreener automatically indexes tokens immediately upon the creation of the liquidity pool and the initial transaction activity~\cite{dexscreener_listing}, while CoinSniper requires only a basic application process, after which the projects appear immediately on their New Listings Page without further verification~\cite{coinsniper_listing}. 
Our scrapers collected 30,771 tokens from DexScreener and 44,611 from CoinSniper, resulting in 65,021 unique tokens.

Although these platforms are commonly used to monitor meme coins due to their focus on newly launched tokens on DEXs, they could list a wide range of token types. To distinguish meme coins from other digital assets, we strictly adhere to the formal definition established by the SEC, which characterizes them as "crypto assets inspired by internet memes, characters, current events, or trends"~\cite{meme_coins_sec}.
Our analysis revealed that these cultural and thematic associations are consistently embedded within token names, creating distinctive linguistic patterns that can be systematically identified. Using this insight, we developed a name-based classification approach to detect meme coins within our larger dataset.

\mypara{Name-Based Classification.}
We begin by analyzing the names of verified meme coins collected in Sec.~\ref{subsubsec:know_meme_coins} to identify common keywords that characterize meme tokens. To extract these keywords, we apply Term Frequency–Inverse Document Frequency (TF-IDF) analysis to our verified meme token dataset. This technique identifies distinctive terms by assigning higher weights to words that appear frequently in meme coin names but are uncommon in general cryptocurrency naming~\cite{ramos2003using}.
Our methodology proceeds as follows: First, we preprocess token names by converting them to lowercase and removing special characters. Next, we tokenize the names into individual words (unigrams) and compute their TF-IDF scores. To determine the optimal number of keywords to retain, we apply the elbow method, which identifies the point where additional keywords provide diminishing marginal value~\cite{bholowalia2014ebk}.
As shown in Fig.~\ref{fig:tf_idf_names} in the Appendix, the elbow analysis suggests 150 keywords as the optimal cutoff point. We then manually review these candidates to remove general cryptocurrency terms (such as SOL or ETH) that appear due to cross-chain versions of the same token (\eg, Dogecoin (SOL)). This refinement yields 126 meme-specific keywords.
The resulting keyword set reflects the market's thematic dominance: Animal-related terms (\textit{dog}: 7.24\%, \textit{cat}: 6.92\%, \textit{inu}: 4.82\%) and trend-based keywords (\textit{AI}: 4.96\%) are heavily represented.
We apply this name-based classification to the token dataset collected in Sec.~\ref{subsubsec:new_dex_meme_coins}, identifying tokens as meme coins if their names contain one or more of the 126 meme-specific keywords. This approach detects 26,557 meme coins.

\mypara{Pump.fun-Based Classification.}
In addition to keyword matching, we directly include tokens originating from the \textit{pump.fun} platform (identifiable by the .pump suffix in their addresses). Two factors justify this platform-specific inclusion. First, \textit{pump.fun} explicitly markets itself as a specialized platform for meme coin creation~\cite{pump_fun_platform}, establishing a high baseline probability that its listed assets are meme-related. Second, to empirically validate this assumption, we checked manually 100 randomly sampled pump' tokens. The inspection confirmed that 100\% of the sampled tokens featured names referencing internet memes, pop culture, or viral trends, thereby strictly aligning with the SEC's definition~\cite{meme_coins_sec}. In particular, this manual verification revealed also niche memes not captured by the name-based classification. In this way, we mitigate false negatives and ensure the inclusion of emerging trends.
With this approach, we identified 4,140 meme coins, detecting overall 28,975 meme coins.

By combining the known meme coins collected in Sec.~\ref{subsubsec:know_meme_coins} with the new tokens, we obtain 35,220 meme coins.

\subsubsection{Dataset Refinement}
\label{subsubsec:dataset_refinement}
Manual inspection of high market cap or high-priced tokens reveals some false positives, such as stablecoins or wrapped tokens. 
We apply a three-stage filtering process to eliminate misclassified tokens and retain only meme coins, minimizing false positives even at the cost of excluding some legitimate ones. This conservative approach prioritizes precision over recall, enhancing the reliability of our subsequent analyses.

First, we use CoinMarketCap's API~\cite{coinmakretcap_api} to identify and remove stablecoins. By cross-referencing our dataset against CoinMarketCap's list of 82 labeled stablecoins, we discard 4 tokens, reducing our dataset to 35,116 tokens.

In the second refinement stage, we focus on identifying non-meme tokens with significant market presence. 
Specifically, we examine tokens that meet either of two conditions: a price above \$0.80—since meme coins typically have very low prices~\cite{nani2022doge}, this threshold helps capture stablecoins that may be slightly depegged—or a market capitalization exceeding $10^7$, as the cumulative distribution function (CDF) in Fig.~\ref{fig:cdf_dataset_refinement} (Appendix) shows that most tokens in our dataset fall below this value.
This process identifies 525 tokens for detailed review. We then use CoinGecko’s API~\cite{coingecko_api} to retrieve their associated categories and remove those labeled with non-meme classifications (e.g., Staking, Bridged). This step results in the removal of 46 additional tokens.

Finally, we apply a conservative string-matching filter to remove tokens with names containing terms typically associated with non-meme assets: specifically \textit{usd}, \textit{wrapped}, and \textit{staked}. This step identifies and excludes 182 tokens, resulting in a final dataset of 34,988 meme coins, which we publicly release~\cite{meme_coin_repository}. 
Tab.~\ref{tab:meme_coins_summary} provides an overview of the tokens collected and the filtering steps applied to refine the dataset.

\begin{table}[ht]
    \small
    \centering
    \caption{Summary of meme coin data collection.}
    \BeginAccSupp{ActualText={This table provides a summary of the meme coin data collection and filtering process.
    It details the number of tokens collected from various sources like CoinMarketCap (2,930), CoinGecko (2,107), and GeckoTerminal (4,735), totaling 8,852 known meme coins.
    Additional token sources include CoinSniper (44,611) and DexScreener (30,771), with a subtotal of 65,021 other tokens.
    Meme coins detected through classification methods include Name-Based Classification (26,557) and Pump.fun Classification (4,140), totaling 28,975 detected meme coins.
    The total meme coins before filtering amount to 35,220.
    The table also lists filtering steps, including the removal of 4 stablecoins, 46 non-meme tags, and 182 wrapped/USD/staked tokens.
    The final meme coin dataset after filtering contains 34,988 tokens.}}%
    \EndAccSupp{}
    \begin{tabular}{l r}
        \toprule
        \textbf{Source} & \textbf{Tokens Collected} \\
        \midrule
        CoinMarketCap & 2,930 \\
        CoinGecko & 2,107 \\
        GeckoTerminal & 4,735 \\
        \textit{Subtotal (Known Meme Coins)} & \textit{8,852} \\
        \midrule
        CoinSniper & 44,611 \\
        DexScreener & 30,771 \\
        \textit{Subtotal (Other Tokens)} & \textit{65,021} \\
        \midrule
        Name-Based Classification & 26,557 \\
        Pump.fun Classification & 4,140 \\
        \textit{Subtotal (Detected Meme Coins)} & \textit{28,975} \\
        \midrule
        \textbf{Total Meme Coins Before Filtering} & 35,220 \\
        \midrule
        Removed Stablecoins (CoinMarketCap) & 4 \\
        Removed Non-Meme Tags (CoinGecko) & 46 \\
        Removed Tokens (Wrapped/USD/Staked) & 182 \\
        \midrule
        \textbf{Final Meme Coin Dataset} & \textbf{34,988} \\
        \bottomrule
    \end{tabular}
    \label{tab:meme_coins_summary}
\end{table}

\subsection{Dataset Overview}
\label{subsec:dataset_overview}
We collected the data from all targeted aggregators in mid-October 2024. For each of the 34,988 tokens in our dataset, we store the token name, symbol, contract address, and blockchain. In particular, the contract address and blockchain serve as unique identifiers, enabling the tracking of tokens across different platforms.

Concerning data coverage and limitations, we note that tokens sourced from CoinMarketCap and CoinGecko include aggregated data from both Centralized (CEX) and Decentralized Exchanges (DEX). For tokens identified solely through DEX aggregators, our view is limited to DEX.
Nevertheless, this does not introduce coverage gaps: If a meme coin is sufficiently established to be listed on a CEX, it is likely indexed by aggregators such as CoinMarketCap. Consequently, tokens found exclusively on DEX aggregators are early-stage assets that have not yet been listed on centralized platforms.



\subsubsection{Blockchain Distribution}
The choice of blockchain significantly impacts the visibility and adoption of a meme coin. The 34,988 meme coins collected reveal clear trends in platform preference. BNB Smart Chain (BSC) hosts the largest share, with 16,405 tokens (46.88\%), followed by Solana with 11,829 (33.80\%), Ethereum with 3,814 (10.90\%), and Base with 828 (2.36\%).

Ethereum~\cite{ethereum_doc} was the first platform to support smart contracts, laying the foundation for decentralized applications. However, its high fees and scalability limitations led to the development of alternatives. One was the creation of new Layer 1 blockchains like BNB Smart Chain (BSC)~\cite{bnbsmartchain}, which is EVM compatible and offers lower transaction costs. Another was the rise of Layer 2 solutions built on Ethereum itself, such as Base~\cite{base_doc}, which improves scalability while remaining part of the Ethereum ecosystem. In contrast, Solana~\cite{solana_doc} is not EVM compatible and takes a different approach, prioritizing low fees and high transaction throughput through its proof-of-history consensus mechanism.  
Together, these four blockchains account for 32,876 meme coins, representing 93.96\% of our dataset. Given their dominance, our subsequent analysis focuses on tokens deployed on these platforms.

\subsubsection{Addresses Validation}
\label{sec:address_validation}
As the next step, we perform a validation process to identify and discard invalid addresses and non-deployed tokens (\eg placeholders such as \textit{0x000comingsoon} or \textit{upcoming}).
For EVM-compatible blockchains (BSC, Ethereum, and Base), we first validate address format: they must be 42 characters long, start with "0x," and contain only hexadecimal characters. This step identifies 791 invalid addresses (739 in BSC, 51 in Ethereum, 1 in Base). Next, for addresses with the correct format, we use the explorers' APIs (Etherscan~\cite{etherscan_api} for Ethereum, BscScan~\cite{bscscan_api} for BSC, and BaseScan~\cite{base_scan} for Base) to check for their creation transactions. We find 254 addresses, 211 in BSC and 43 in Ethereum, that lacked creation transactions and, thereby, are not deployed contracts.
For Solana tokens, we use Solana's RPC API~\cite{solana_rpc} to verify valid mint addresses, identifying 20 invalid addresses.

Tab.~\ref{tab:validation_results} summarizes the validation process and its results. After removing the 1,065 invalid and non-deployed addresses, the final dataset contains 31,811 tokens (96.76\%): 15,455 on BSC, 11,809 on Solana, 3,720 on Ethereum, and 827 on Base.

\begin{table}[h]
\centering
\small
\caption{Summary of token address validation process}
\label{tab:validation_results}
\begin{tabular}{lcccc}
\toprule
\textbf{Platform} & \textbf{Invalid} & \textbf{Undeployed} & \textbf{Removed} & \textbf{Retained} \\
\midrule
BSC       & 739  & 211  & 950  & 15,455 \\
Ethereum  & 51   & 43   & 94   & 3,720  \\
Solana    & -    & 20   & 20   & 11,809 \\
Base      & 1    & 0    & 1    & 827    \\
\midrule
\textbf{Total} & \textbf{791} & \textbf{254} & \textbf{1,065} & \textbf{31,811} \\
\bottomrule
\end{tabular}
\end{table}

\subsubsection{Temporal Evolution of Meme Coin Creation. }
\label{sec:temporal_evolution_of_memecoins}
To analyze the creation patterns of the meme coins, we collect their contract deployment timestamps. 
For EVM-compatible blockchains, we use their explorers' APIs (Etherscan, BscSca, and BaseScan). Concerning Solana, we leverage the API offered by Alchemy~\cite{alchemy_api}, a blockchain development service that provides tools for building decentralized applications.

\begin{figure}
    \centerline{%
        \BeginAccSupp{ActualText={A detailed plot illustrating the creation dates of meme coins over time, showing temporal trends. It also displays their distribution across different blockchain platforms like Ethereum, Solana, and Binance Smart Chain, indicating which platforms host newer or more numerous meme coins.}}%
            \includegraphics[width=0.49\textwidth]{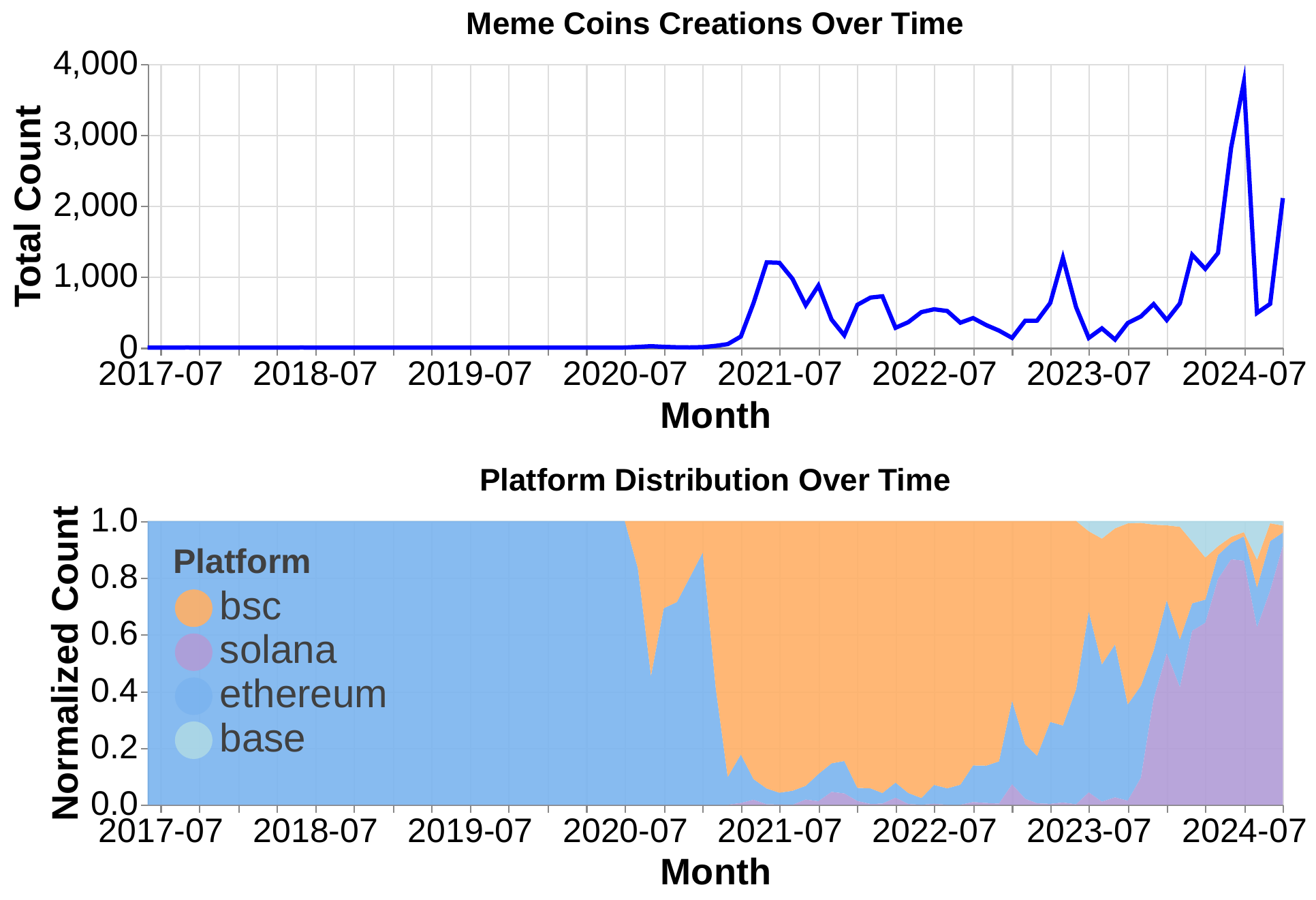}%
        \EndAccSupp{}%
    }
    \caption{Creation date of meme coins over time and their distribution on different blockchains.}
    \label{fig:creation_date_analysis}
\end{figure}

Fig.~\ref{fig:creation_date_analysis} shows the temporal distribution of meme coin creation across different blockchains from 2017 to 2024. The data highlights how platform preferences for meme coin deployment have shifted over time, from Ethereum’s early lead to BSC’s rise in popularity and, more recently, to the growing adoption of Solana and Base. During the initial period (2017–2020), meme coin activity remained low, with fewer than 10 new tokens created per month. This scenario changed significantly due to several key events.

\mypara{Late 2020 Bull Run}. The cryptocurrency bull market, marked by Bitcoin’s rise to over \$60,000~\cite{bull_run_2021}, coincided with two major events: the launch of SHIBA INU and the introduction of Binance Smart Chain (BSC), that significantly boosted meme coin creation thanks to its low transaction fees~\cite{bsc_success}.

\mypara{Pepe Coin \& BONK}. In April 2023, the launch of Pepe coin on Ethereum challenged BSC’s dominance. This was followed by the surge of BONK in October 2023, driven by the introduction of single-sided staking pools~\cite{bonk_staking_pool}.

\mypara{Pump.fun}. The peak in meme coin creation came in the summer of 2024, especially in June and July, when Solana emerged as the leading platform. This surge is largely due to the rapid rise in popularity of pump.fun~\cite{pump_success}.

\section{Tokenomics and Price Growth}
In this section, we group meme coins by their creation dates and analyze their economic characteristics across different blockchains to uncover distinctive patterns within the meme coin ecosystem. We then examine their price movements over three months to understand how their value changes over time.

\subsection{Tokenomic Anatomy Across Blockchains}
\label{subsec:tokenomics}

\begin{figure}[!ht]
  \centering
  \BeginAccSupp{ActualText={A figure containing three CDF plots comparing meme coins across Ethereum, BSC, Solana, and Base. The first plot shows the distribution of price, the second shows the distribution of market cap.}}%
  \EndAccSupp{}
  \subfigure[Price]{%
  \includegraphics[width=.45\textwidth]{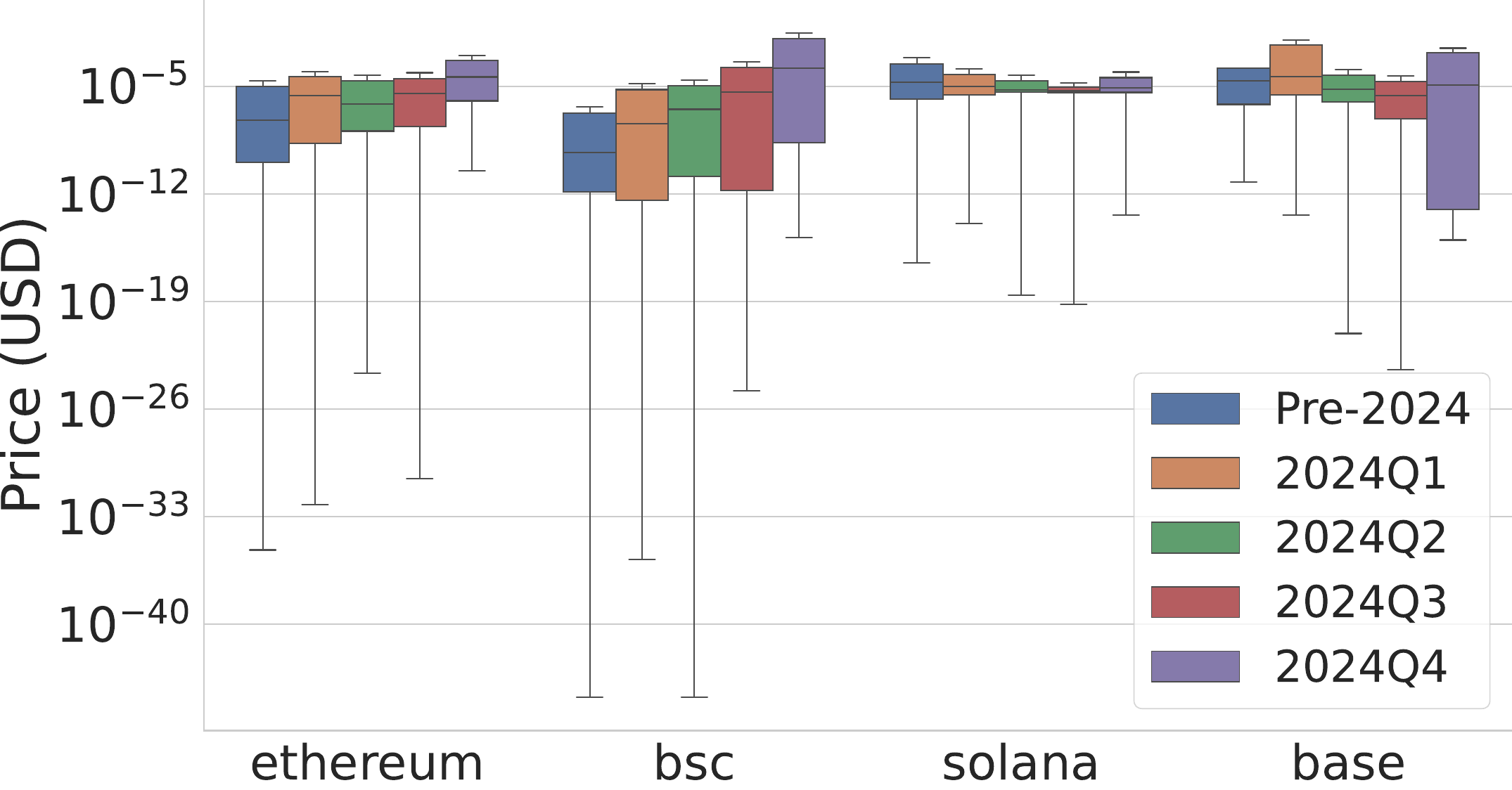}
  \label{fig:price_boxplot}}
  \subfigure[Market Cap]{%
  \includegraphics[width=.45\textwidth]{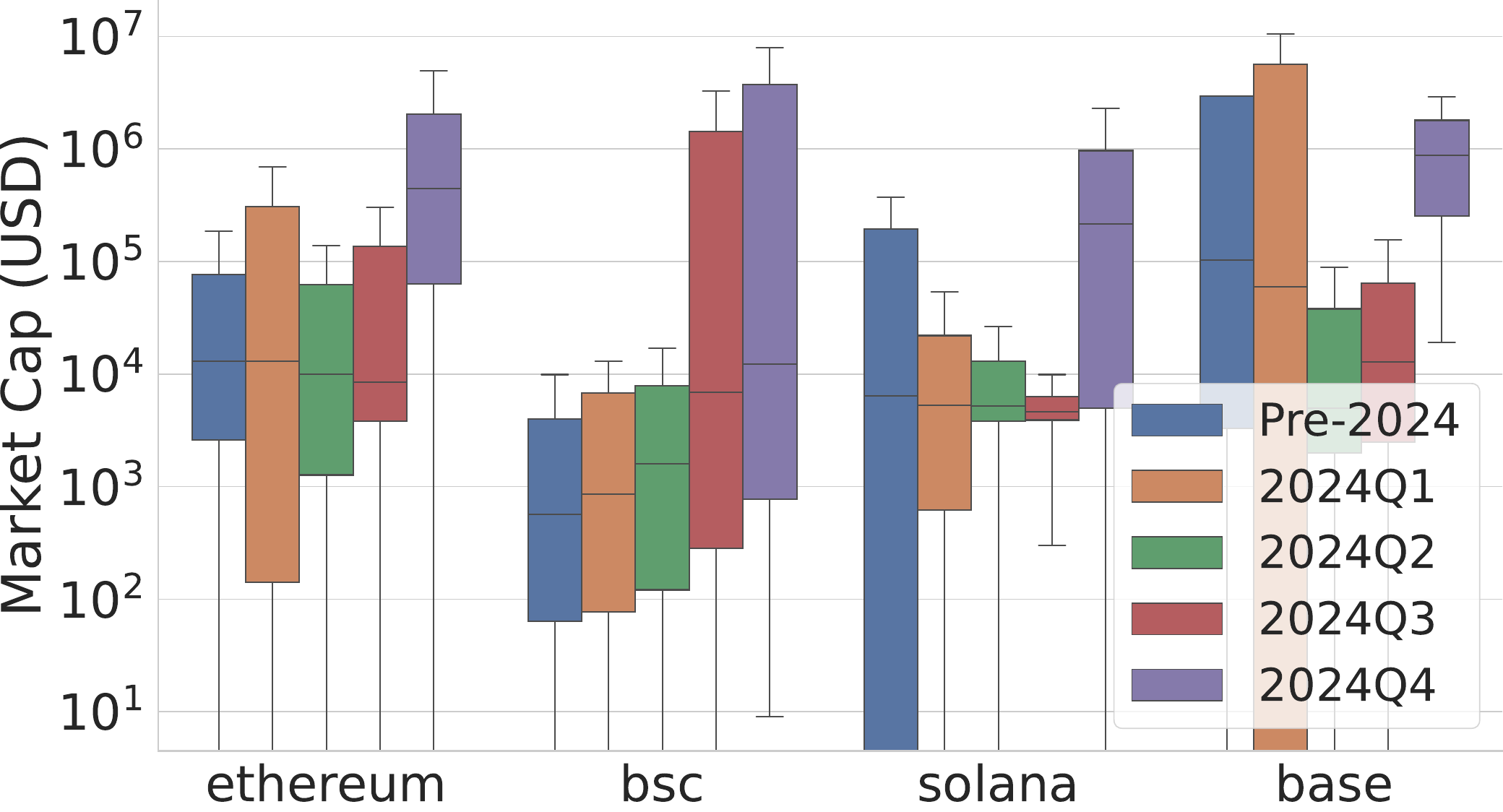}
  \label{fig:marketcap_boxplot}}
  \caption{Distribution of the prices (\ref{fig:price_boxplot}) and market cap (\ref{fig:marketcap_boxplot}) of the meme coins. Ethereum and BSC show sequential growth, with recent tokens ($Q4_{2024}$) reaching valuations orders of magnitude higher than pre-2024 baselines. Solana maintains stable pricing but shares the trend of significantly higher market caps for new launches, while Base exhibits high variability without clear temporal patterns.} 
  \label{fig:box_plots_tokenomics}
\end{figure}

Of the 31,811 meme coins collected, we successfully obtained economic information for 20,692 meme coins (65.05\%) in US dollars during our data snapshot in mid-October 2024. 
The remaining 11,119 tokens for which we cannot retrieve additional data have been collected on DexScreener, Coinsniper, and CoinGecko. 
These platforms derive token price, circulating supply, and market capitalization from liquidity pools where tokens are traded.
However, if a token lacks a liquidity pool at the time of data collection, its price and market cap cannot be determined. This explains why no such information is available for these 11,119 tokens.   
After removing four tokens with a price of zero (all on BSC), the final dataset consists of 20,688 meme coins: 10,468 on Solana, 7,493 on BSC, 2,107 on Ethereum, and 620 on Base.

We group the meme coins in our dataset according to when they were created to examine how their age correlates with their economic characteristics. Specifically, we group tokens into quarterly periods for 2024 ($Q1_{2024}$, $Q2_{2024}$, $Q3_{2024}$, and $Q4_{2024}$), with an additional category for tokens created before 2024 (pre-2024). 
To validate the observed market behaviors, we performed a non-parametric Kruskal-Wallis H test~\cite{kruskal1952use}, followed by Dunn’s post-hoc analysis with Bonferroni correction. We consider differences to be statistically significant only when $p < 0.01$~\cite{dunn1961multiple}. Fig.~\ref{fig:box_plots_tokenomics} illustrates how economic metrics vary across these periods for each blockchain.

\subsubsection{Price Structures}
\label{subsub:price_distribution}

The price analysis reveals an ecosystem characterized by tokens with very low prices, alongside a few high-priced outliers. The overall distribution shows a median price of \$$4 \times 10^{-6}$ with approximately 85\% of tokens priced below \$0.00001. This pricing strategy creates a psychological illusion of accessibility: investors can acquire billions of tokens for minimal capital, fostering the narrative that these assets could replicate the trajectories of Dogecoin or Shiba Inu~\cite{nani2022doge}.

\mypara{Cases of Suspiciously High Pricing}: We identify 14 meme coins with prices exceeding \$100, each employing one of two distinct price strategies. The first involves artificial scarcity mechanisms using regular token burning, as seen in tokens like \textit{Ponzio The Cat} (valued at approximately \$20 billion) and \textit{HayCoin} (\$250,000). 
These tokens have been flagged as scams by security platforms such as Token Sniffer~\cite{token_sniffer} and show systematic supply reduction designed to inflate prices, rewarding early investors at the expense of those who enter later.
The second approach involves the use of single-token supply models, as seen in tokens such as \textit{Potato} (valued at \$55,000). This extreme scarcity is leveraged to justify their unusually high valuations, despite limited utility.

Concerning the temporal analysis of the price, Fig.~\ref{fig:price_boxplot}) reveals that
Ethereum and BSC exhibit sequential price increases across time periods, suggesting a market preference for newly launched tokens. 
Base shows high price variability without statistically significant temporal patterns.
Conversely, Solana demonstrates price consistency across all creation periods, maintaining a median of approximately $8 \times 10^{-6}$ regardless of token age.

\subsubsection{Market Capitalization}
\label{subsec:market_cap_analysys}
Market capitalization, calculated by multiplying the circulating supply of a token by its current price, is a key indicator of the market value of a project. Among the meme coins we analyzed, the median market cap is just \$4,100, highlighting their low market cap and highly speculative nature.

\mypara{Cases of High Market Caps with Low Liquidity}: A few meme coin projects report extremely high market cap. While some well-known tokens, such as Dogecoin (\$53 billion), Shiba Inu (\$15 billion), and Pepe (\$8 billion), have substantial and legitimate market value, these are rare exceptions in a space often marked by value manipulation. For example, \textit{catgirl a} claims an implausible market cap of \$151 billion despite having almost no liquidity, suggesting how unreliable market capitalization can be as a standalone metric for meme coins.
Upon investigating the 47 meme coins reporting market caps above \$1 billion, we detect significant evidence of market manipulation. More than half have liquidity under \$1,000, and 88.1\% show liquidity below 0.1\% of their reported market cap. This discrepancy points to a common tactic: inflating token prices through small, strategic trades in low-liquidity environments, a form of manipulation detailed in Sec.~\ref{subsec:price_inflation}.

Regarding the cross-chain analysis (Fig.~\ref{fig:marketcap_boxplot}), we found Solana and Ethereum show statistically significant increases in market cap for $Q4_2024$ tokens, with Ethereum tokens reaching a median of \$443K and Solana tokens \$214K. BSC shows a similar but weaker trend with $Q4_2024$ tokens at \$12K, while Base exhibits high variability without consistent temporal patterns.
This strong preference for newly launched tokens is more consistent across blockchains than pricing patterns, suggesting that token novelty may be a stronger driver for investors than the underlying blockchain infrastructure.

\subsubsection{Cross-Platform Economic Patterns}
\label{subsec:blockchain_economic_signatures}
Our analysis of price, supply, and market capitalization reveals that each blockchain has its own economic pattern. 
Ethereum meme coins show higher prices for newer tokens, extremely large token supplies, and high market caps. BSC presents a similar pattern to Ethereum but with key differences. While it shares hyperinflated token supplies and rising price trends for newly created tokens, BSC meme coins maintain consistently lower market caps than all other platforms. 
Conversely, Solana demonstrates remarkably consistent pricing regardless of token age, standardized supply structures, and moderate market caps, reflecting its more structured and possibly regulated approach to meme coin creation. 
Instead, Base exhibits characteristics of an emerging ecosystem with cyclical price volatility rather than consistent trends, moderate supply metrics, and dramatic variations in market capitalization.

\subsection{Price Performance Evaluation}
\label{sec:performance_analysis}
To study the price evolution of meme coins over three-months following our initial snapshot, we conducted a follow-up data collection in mid-January 2025.

\subsubsection{Data Availability and Retrieval}
\label{subsubsec:data_collection_results_performance}
Of the 20,688 meme coins we analyzed in Sec.\ref{subsec:tokenomics}, we successfully obtained updated price data for 20,327 tokens (98.25\% of the original sample), including 10,438 on Solana, 7,355 on BSC, 1,922 on Ethereum, and 612 on Base.

Concerning the 361 tokens with missing price, we originally collected them from multiple sources (with some overlap): 12 from Gecko Terminal, 350 from DexScreener, and 21 from Coingecko.
As discussed in Sec.~\ref{subsub:price_distribution}, Gecko Terminal and DexScreener require tokens to maintain active liquidity pools to determine prices. Thus, the absence of price data for these tokens likely indicates their liquidity pools were removed during our study period. Meanwhile, Coingecko may delist tokens failing to meet its listing criteria~\cite{coingecko_listing}, particularly those with poor liquidity. 
The blockchain with the highest percentage of tokens lacking updated price is Ethereum with 185 tokens (8.78\%), followed by BSC with 138 (1.84\%), Base with 8 (1.29\%), and Solana with 30 (0.29\%).

\subsubsection{Price Change Distribution}
\label{subsec:return_price_performance}
\begin{figure}
    \centerline{\includegraphics[width=0.49\textwidth]{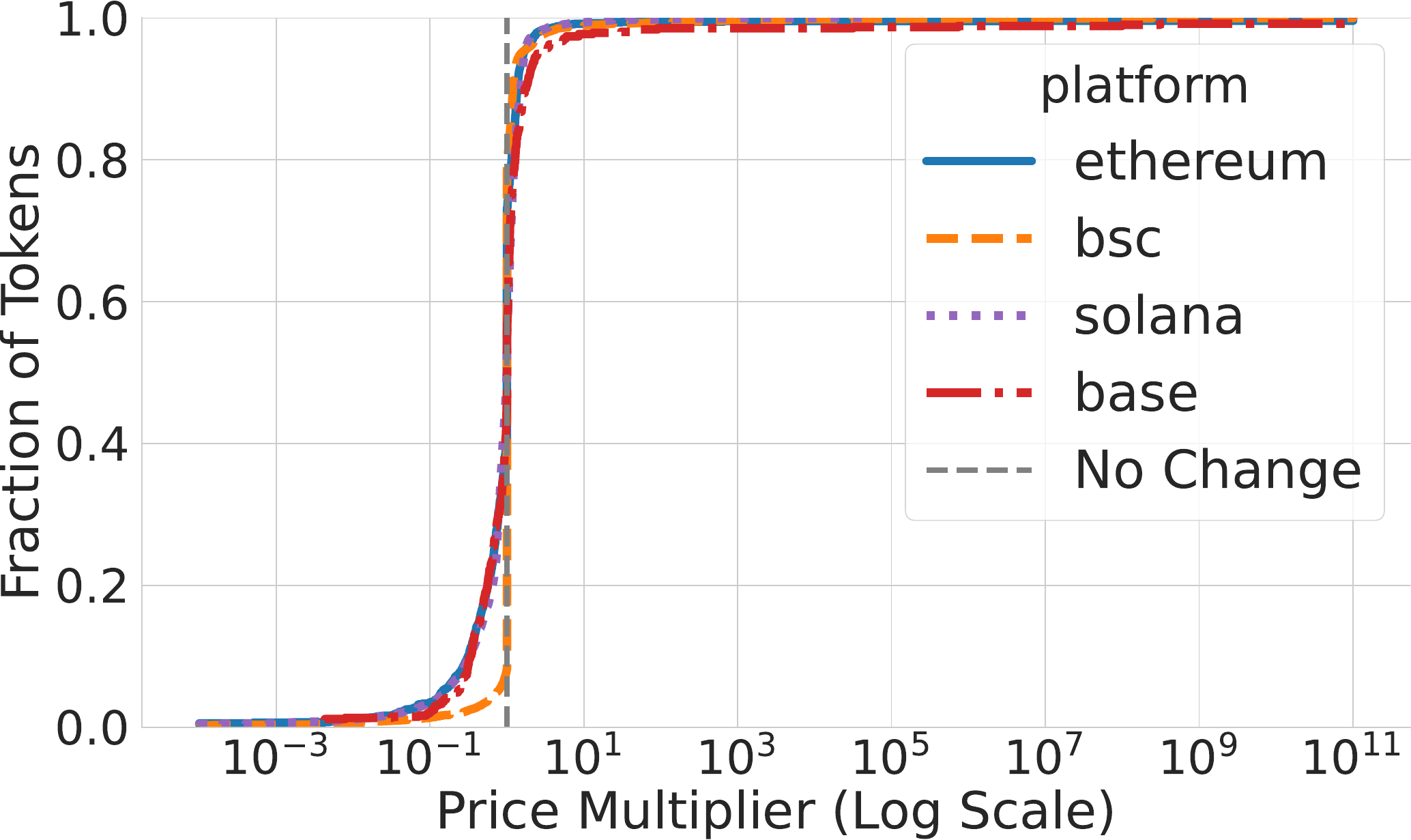}}
    \caption{Log-CDFs of meme coin price changes by blockchain over three months. Many tokens experience almost no price change, suggesting they are likely inactive.}
    \BeginAccSupp{ActualText={A plot showing the CDFs of the price changes of meme coins grouped by different blockchains (Ethereum, BSC, Solana, and Base).}}%
    \EndAccSupp{}
  \label{fig:cdf_prices_evolution}
\end{figure}

Fig.~\ref{fig:cdf_prices_evolution} presents the cumulative distribution functions (CDFs) of meme coin price performance across different blockchains. To visualize the extreme disparity between losses and gains, we use a logarithmic scale for the price multiplier, defined as $Price_{final} / Price_{initial}$.
Specifically, 6,938 tokens (34.13\%) experience price decreases, 6,462 tokens (31.79\%) maintain stable prices, and 6,927 tokens (34.07\%) see price increases.

\mypara{Negative Return Tokens.}
The 6,938 tokens that decreased in value lost an average of 38.42\%, with a median loss of 27.97\%. Substantial losses are common across the dataset: 946 tokens (13.63\%) lost more than 80\% of their value, while 2,299 tokens (33.13\%) lost more than 50\% of their value.
In particular, performance varies substantially across blockchains. Ethereum shows the largest median price decrease at 42.13\%, followed by BSC and Base with median decreases of 38.19\% and 36.95\%, respectively, while Solana tokens demonstrated greater resilience with a median decrease of only 26\%.
However, when analyzing the proportion of tokens with price losses within each blockchain, Solana presents the highest rate (50.43\% of its tokens), followed by Base (46.41\%), Ethereum (40.95\%), and BSC with the lowest rate (8.21\%).

\mypara{Inactive Tokens.}
Interestingly, 6,462 tokens (31.79\% of the sample) maintained exactly the same price over the three-month period. This price stability appears to result from complete inactivity rather than market equilibrium, as all of these tokens recorded zero trading volume during our observation period.
Unlike the 361 tokens for which price updates were unavailable, services like DexScreener continue recording the prices of these inactive coins as they maintain open liquidity pools.
Regarding the distribution of inactive tokens across blockchains, the majority are on BSC, with 5,277 tokens (71.75\% of BSC tokens with updated price). Ethereum follows with 609 inactive tokens (31.69\%), then Solana with 523 tokens (5.01\%), and Base with 53 tokens (8.66\%).

\mypara{Positive Return Tokens.}
Among the 6,927 tokens that increased in value, extreme outliers skewed the mean increase to $10^{35}$\%, while the median is only 21.46\%.
The distribution of positive returns shows that 707 tokens (10.21\%) gained more than 100\%, and 171 tokens (2.47\%) achieved returns exceeding 1,000\%.
Across blockchains, Ethereum tokens present the highest median price increase at 29.33\%, while tokens on other chains averaged approximately 20\% growth. The proportion of tokens with positive price performance is highest on Base and Solana (44\% each), followed by Ethereum (25\%), and lowest on BSC (19.65\%).

If we exclude inactive tokens (those with a stable price), Fig.~\ref{fig:cdf_prices_evolution} suggests that investing in meme coins could result in positive growth, sometimes yielding significant profits.  
However, this analysis overlooks two key factors. First, it does not account for the 361 tokens for which updated price data was unavailable, many of which likely lost all their value. Second, as discussed in Sec.~\ref{subsec:market_cap_analysys}, meme coins with high returns may be subject to market manipulations that artificially inflate their perceived value. To explore this further, we examine the 707 tokens that exhibited exceptionally high returns (>100\%) after three months.

\section{Artificial Growth Strategies}
\label{sec:growth_analysis}
This section analyzes high-performing meme tokens: First, anomalous activity patterns; then, evidence of manipulations designed to distort the market.
In particular, in order to analyze potential manipulations, we collected OHLCV (Open, High, Low, Close, Volume) data with daily granularity from our initial data collection in mid-October 2024 through mid-January 2025 using the Gecko Terminal APIs~\cite{gecko_terminal_api}.

\subsection{Market Anomalies in High-Return Tokens}
\label{sec:anomalies} 
To better understand the behavior of the 707 high-performing tokens, we investigate anomalies in their market characteristics. 
To identify these irregularities, we leverage external security APIs (\eg GoPlus~\cite{go_plus}, De.Fi~\cite{defi_scanner}, and Soul Scanner~\cite{soul_scanner}) which aggregate risk signals such as hidden ownership concentration.
This investigation serves as a foundational risk assessment. While the detected anomalies are strongly correlated with manipulative behavior, they represent warning signals rather than definitive proof of manipulation. We establish concrete evidence of manipulation through the detailed analyses presented in Sec.~\ref{subsec:washtrading} and Sec.~\ref{subsec:price_inflation}.

\subsubsection{Ownership Concentration Anomalies}
\label{subsubsec:ownership_concentration}
Ownership concentration occurs when a small number of wallets control most of the token supply. This concentration poses significant risks for investors, as major holders can manipulate prices through coordinated trading activities at the expense of smaller investors who often lack visibility into these ownership structures. 
We detect severe ownership anomalies across many high-return meme coins, with various techniques employed sometimes to concentrate token control while creating an illusion of distributed ownership.

\mypara{Concentrated Holdings.} 
We identify 385 tokens (54.46\%) with an alarming concentration of ownership among a small number of holders. The top 10 holders controlled an average of 77.85\% of these tokens, with a median of 87.01\%. 
Following standards used by established blockchain security platforms such as GoPlus~\cite{go_plus} and CertiK~\cite{certik_scan}, we identify problematic concentration when the top holders control more than 30\% of a token's supply. This is the case for 359 tokens (50.78\% of high-return tokens).

\mypara{Bundle Buy Concealment.}
A strategy developers can use to obscure their token holdings is leveraging bundle buys (e.g., JITO bundles~\cite{jito_bundle}), an approach that allows them to group multiple transactions together and execute them in a specific order. This enables the creation of bundles that first add liquidity (depositing tokens into a liquidity pool) and then immediately swap tokens, ensuring trades are executed right after liquidity is added. By distributing the swapped tokens across multiple wallets, they can mask true ownership and create the appearance of distributed holdings.
We observe 133 tokens (18.81\%) utilizing bundle buys, with an average of 15.70\% of their supply involved. As for the concentrated holdings, we consider a bundle buy an anomaly when it exceeded 30\% of a token's supply, which occurred in 20 cases (2.83\%). 

\mypara{Fresh Address Distribution.}
Another concealment technique is the use of multiple newly created addresses.
This tactic allows creators and top holders to give the illusion of decentralized ownership while maintaining control for coordinated manipulation.  
We detect 8 projects where a significant portion of tokens (more than 30\%) are held in fresh addresses, wallets with no prior blockchain history. On average, these addresses control 40.25\% of the token supply, with a median of 38\%.

\mypara{Strategic Airdrop Distribution.}
Airdrops involve distributing tokens for free to multiple addresses and are commonly used for legitimate promotional purposes. However, they can also serve as a concentration tactic. By strategically distributing tokens to multiple controlled wallets, token creators can create the illusion of decentralized ownership while maintaining effective control. 
The average percentage of tokens distributed via airdrops is 63.43\%, with 24 tokens (3.40\%) having more than 30\% of their supply allocated by airdrops.

\subsubsection{Honeypots}
Unlike ownership concentration anomalies, which focus on token distribution, honeypots represent a trading activity anomaly. These fraudulent tokens appear legitimate but are designed with malicious code that prevents investors from selling after purchase, effectively trapping their funds. This deceptive structure allows creators to collect investment capital while ensuring investors cannot exit their positions, regardless of how much the token's value drops. We find 28 meme coins flagged as honeypots by Go Plus Token Security~\cite{go_plus}, representing a significant risk to unwary investors seeking high returns in the meme coin market.

\subsubsection{Anomaly Prevalence}
Tab.~\ref{tab:anomaly_stats} provides a summary of the statistics of the anomalies detected. 
Of the 707 high-performing tokens, 412 meme coins (58.27\%) exhibit at least one anomaly indicator. 
The most common issue is ownership concentration, where the top 10 holders own over 30\% of the supply, occurring in 87.14\% of the detected anomalies. Concerning the other indicators, although they are less frequent, they still help in flagging suspicious tokens that might otherwise escape detection.
\begin{table}[ht]
    \small
    \centering
    \caption{Anomaly statistics for high-performing meme coins.}
    \begin{tabular}{l c c}
        \toprule
        \textbf{Anomaly Type} & \textbf{\# tokens} & \textbf{Percentage in anomalies} \\
        \midrule
        Top Holders & 359 (50.78\%) & 87.14\% \\
        Honeypots & 28 (3.96\%) & 6.80\% \\
        Airdrop & 24 (3.39\%) & 5.83\% \\
        Bundle Buys & 20 (2.83\%) & 4.85\% \\
        Fresh Addresses & 8 (1.13) & 1.94\% \\
        \midrule
        \textbf{Total tokens} & 412 (58.27\%) & 100\%\\
        \bottomrule
    \end{tabular}
    \BeginAccSupp{ActualText={The table shows the anomaly statistics for meme tokens with high returns, including the number and percentage of tokens affected by each type of anomaly.}}%
    \EndAccSupp{}
    \label{tab:anomaly_stats}
\end{table}
The prevalence of anomalies in high-performing tokens raises serious doubts about their market success. To establish more concrete proof of artificial market activity, the following two subsections examine manipulation techniques that directly impact token valuation and visibility. 

\subsection{Wash Trading Analysis and Detection}
\label{subsec:washtrading}
Wash trading is a deceptive practice in which the same entities buy and sell tokens to create an illusion of market activity. This manipulation inflates the trading volume without real ownership changes or market risk, creating a false perception of popularity and demand~\cite{le2021wash}.
Meme coins are especially vulnerable to this strategy, as their success often depends on social sentiment and perceived popularity~\cite{nani2022doge}. 


\subsubsection{Initial Detection Methodology}
To detect potential wash trading, we identify suspicious trading patterns where volume increased dramatically (>500\% compared to the previous day) while the price remained nearly unchanged (absolute price change <5\%). This pattern contrasts with natural market behavior, where significant volume surges typically drive notable price movements. These thresholds are heuristics implementing definitions provided in prior works~\cite{le2021wash,victor2021detecting}.
The 500\% volume threshold focuses on extreme trading activity that deviates from normal market patterns, while the 5\% price stability criterion helps distinguish these suspicious activities from legitimate high-volume trading that would typically cause significant price changes.
Applying this method to the OHLCV data, we detect 3,266 potential wash trading instances involving 379 tokens. 

\subsubsection{Advanced Detection Techniques}
\label{subsec:advanced_detection_wash_trading}
To strengthen our wash trading detection and provide more concrete evidence, we use the Dune Analytics API~\cite{dune_api} to collect detailed transaction data. This enables us to identify the traders (makers) involved in suspicious transactions on the days flagged by our initial methodology. We apply three different techniques to detect wash trading activities.

\mypara{Zero-Risk Position.}
Wash trading, by definition, is a zero-risk manipulation where traders buy and sell without taking actual market risk. 
Following the methodology established by~\cite{la2023game}, we look for makers who bought and sold almost identical amounts of the same token on the same day. To account for slippage and transaction fees, we allow for a small 2\% difference between buy and sell volumes, consistent with methodologies used in~\cite{la2023game}.
This approach detects 989 manipulations (30.28\% of suspected cases), 276 involved tokens (72.82\% of suspected tokens), and 2,687 unique makers.

\mypara{Circular Volume.}
A strong indicator of wash trading is when nearly all trading volume comes from the same set of addresses acting as both buyers and sellers. Thus, inspired by the cyclical pattern analysis in~\cite{victor2021detecting}, we analyze instances where at least 99\% of daily trading volume is generated solely by makers who both bought and sold the token on the same day.
This technique identifies 223 (6.83\%) suspicious activities involving 131 (34.56\%) tokens and 1,148 unique makers engaged in circular trading.

\mypara{Coordinated Manipulator Network.}
To uncover potential collusion networks, we examine whether the same makers participated in multiple wash trading events for the same token, focusing specifically on identifying coordinated manipulators who appeared in every detected wash trading day for a given token. This technique detects 42 manipulations involving 17 tokens and 178 unique makers.

Tab.~\ref{tab:wash_trading_results} summarizes the wash trading detection analyses.
By combining the three approaches, we identify 1,043 events presenting strong evidence of wash trading (31.94\% of suspected instances) and 287 tokens affected (75.73\% of suspected tokens).
The blockchain distribution of the manipulated tokens reveals that most are on Solana, with 183 tokens, followed by BSC with 52, Ethereum with 29, and Base with 23.

\begin{table}[h]
\centering
\small
\caption{Wash trading detection results by methodology.}
\label{tab:wash_trading_results}
\begin{tabular}{lccc}
\toprule
\textbf{Analysis Method} & \textbf{Events Detected} & \textbf{Tokens Affected} \\
\midrule
Zero-Risk Position & 989 (30.28\%) & 276 (72.82\%) \\
Circular Volume & 223 (6.83\%) & 131 (34.56\%) \\
Coord. Manipulators & 42 (1.29\%) & 17 (4.49\%) \\
\midrule
\textbf{Total} & \textbf{1,043} & \textbf{287} \\
\bottomrule
\end{tabular}
\end{table}

The overlap of the techniques reveals that most wash trading activities (74.11\%) are uniquely identified by zero-risk position, while only a smaller proportion is exclusively detected through circular volume (2.78\%). 
At the token level, the zero-risk position detects 96.17\% of the coins suspected of wash trading, suggesting this approach is the foundation of most wash trading schemes.

\subsubsection{Wash Trading Characteristics}
After detecting instances of wash trading, we analyze their patterns to understand how wash traders operate.

\mypara{Recurrent Manipulations.}
Affected tokens experience an average of 3.63 wash trading operations, with a median of 3 and a maximum of 23, a result consistent across all blockchains.
This pattern suggests recurring manipulation rather than isolated activities. Multiple wash trading operations on the same token likely serve to sustain the illusion of legitimacy and create a perception of growing interest over time.

\mypara{Extreme Volume Increases.}
Wash trading operations create dramatic spikes in trading volume compared to the previous day, showing a median increase of 1,772\% and an average increase distorted by extreme outliers reaching $10^8$\%. During wash trading events, the average absolute trading volume is \$10,385, with a median of \$859, presenting platform-specific differences. BSC shows the lowest median volume (\$394), followed by Base (\$553) and Solana (\$964), while Ethereum records the highest median volume (\$1673). This pattern likely reflects Ethereum's higher transaction fees, which make small-volume wash trades economically impractical compared to other platforms with lower gas costs.

\mypara{Temporal Patterns and Lifecycle Analysis.}
\begin{figure}
    \centerline{\includegraphics[width=0.49\textwidth]{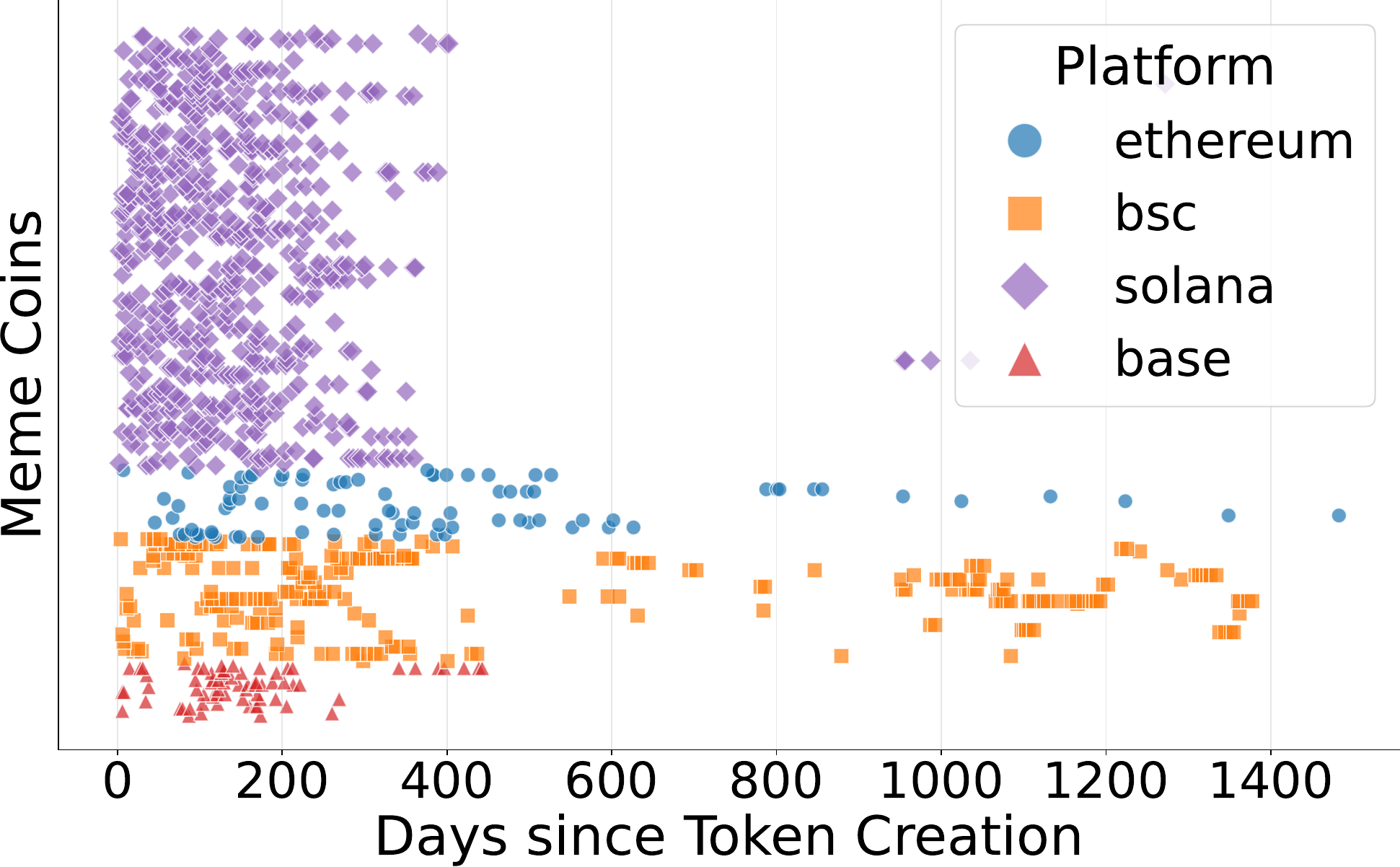}}
    \caption{Temporal distribution of wash trading activity per token, relative to its creation date, across different blockchains.}
    \BeginAccSupp{ActualText={The figure shows a scatterplot of wash trading activities occurring on each meme coin since its creation date, with different colors representing different blockchains.}}%
    \EndAccSupp{}
  \label{fig:wash_trading_temporal_pattern}
\end{figure}
Fig.~\ref{fig:wash_trading_temporal_pattern} illustrates the timeline of wash trading activity for meme coins. 
Most tokens experience their first manipulation within a median of 85 days, with some targeted earlier: 6.27\% (18) within their first week and 24.27\% (71) within their first month. Specifically, Solana tokens are targeted earlier, with a median of just 61 days after creation, compared to medians of at least 101 days on other blockchains.
Moreover, these wash trading campaigns often continue for extended periods, averaging 86 days, indicating prolonged and sustained manipulation. Only 40.07\% (115) of manipulated tokens experience activity for one month or less.
For tokens subjected to multiple manipulations, the average time between consecutive operations is 33 days, with a median of 16 days. Specifically, 19.94\% (208) of wash trading events occur within at most one week, and 50.24\% (524) occur within at most one month.
This pattern holds across blockchains, except for BSC, which has a slightly shorter median interval of 9 days.

\mypara{Few Wash Traders.}
Wash trading activities typically involve very few participants, with an average of 3.92 actors and 514 operations (49.28\%) conducted by a single actor.
This manipulation pattern suggests coordinated market manipulation carried out by a small group of entities or even individuals, a trend also observed in previous studies on wash trading~\cite{la2023game}.


These repeated wash trading operations performed by a few entities on the same tokens over extended periods suggest a strategic purpose: creating trading volume to maintain the appearance of legitimate market activity. 
Specifically, wash trading can serve two strategic purposes. First, it helps tokens reach the minimum trading volume required for listing on established cryptocurrency aggregators~\cite{coingecko_listing,coinmarketcap_listing}, ensuring visibility to potential investors. Second, it increases the ranking of a meme coin on platforms such as DexScreener~\cite{dexscreener}, which sorts tokens by trading volume. This increased visibility attracts unsuspecting investors, who perceive the artificial volume as evidence of genuine market interest.


\subsection{Liquidity Pool-Based Price Inflation (LPI)}
\label{subsec:price_inflation}

In Sec.~\ref{subsec:market_cap_analysys}, we identified meme coins reporting suspiciously high market cap despite having minimal liquidity. Investigation of these cases reveals a previously undocumented manipulation strategy that directly inflates token prices through exploitation of automated market maker (AMM) mechanics. We term this technique \textit{Liquidity Pool-Based Price Inflation} (LPI) and provide its first formal characterization.

\mypara{Definition.}
We define LPI as a price manipulation technique that exploits the constant product formula ($x \times y = k$) used by AMM-based liquidity pools to generate disproportionate price increases with minimal capital investment. In this formula, $x$ represents the quantity of one token (\eg ETH) and $y$ represents the quantity of another token (\eg a meme coin). The constant $k$ must be maintained across all trades, meaning any purchase that removes tokens from the pool necessarily increases the price of the remaining tokens.
Critically, in low-liquidity pools, even small purchases significantly alter the token ratio, triggering exponential price increases. Manipulators exploit this property to inflate the price with negligible capital, creating a misleading impression of rapid growth that attracts unsuspecting investors.

\mypara{Illustrative Example.}
Consider a liquidity pool containing 0.01 ETH and 100 MEME tokens, where $k = 0.01 \times 100 = 1$. At this state, each MEME token is worth 0.0001 ETH. 
When a manipulator purchases 50 MEME tokens, they must add enough ETH to maintain the constant product. To remove 50 MEME (leaving 50 in the pool), the calculation becomes: $50 \times y = 1$, thus $y = 0.02$ ETH must remain in the pool. Since the pool originally contained 0.01 ETH, the manipulator must add 0.01 ETH to acquire 50 MEME tokens.
After this transaction, the pool contains 50 MEME and 0.02 ETH, establishing a new price of 0.0004 ETH per MEME, a 300\% price increase achieved through a single transaction costing just 0.01 ETH (currently about \$30-40).

\subsubsection{LPI Detection} 
\label{sec:lpi_detection}
To detect LPI activities, we develop a conservative two-phase methodology. The first phase focuses on identifying statistically anomalous trading patterns by applying two main criteria: (1) a significant price increase exceeding 100\% within a 24-hour period, which may indicate artificial price inflation; and (2) either minimal growth in trading volume compared to the previous day ($\leq 20\%$ increase) or an absolute trading volume below \$1,000, both of which distinguish synthetic price movements from legitimate market demand. This approach isolates the defining characteristic of LPI operations, namely, abrupt price increases that lack corresponding trading activity. When applied to our OHLCV dataset, this screening process identifies 871 events involving 149 tokens.

The second phase implements a more stringent verification step to minimize false positives. Using the Dune Analytics API, we retrieve granular transaction-level data for each flagged event and introduce two additional constraints. First, buy volume must account for at least 90\% of the total daily trading volume, indicating sustained buying pressure rather than balanced market activity. Second, the number of unique trading entities must not exceed 10, implying coordinated or strategic behavior rather than organic participation by a broad user base. 
This methodology detects 72 price manipulations affecting 40 distinct meme coins, 12 on Solana, 11 on BSC, 10 on Ethereum, and 7 on Base. 

\subsubsection{LPI Analysis}
Next, we examine LPI events focusing on their financial impact, actor participation, and the targeted tokens.

\begin{figure}
    \centerline{\includegraphics[width=0.49\textwidth]{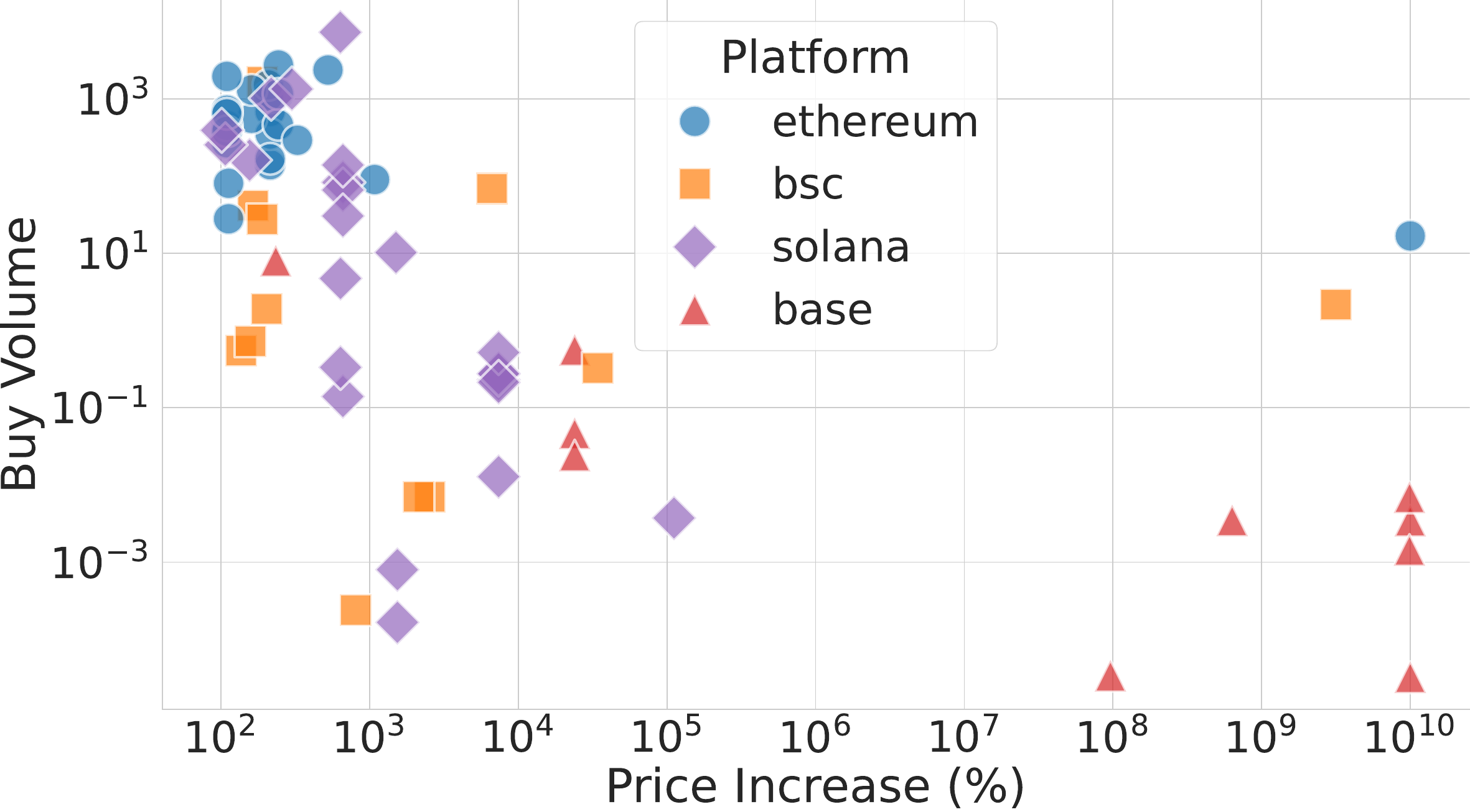}}
    \caption{Relationship between price increase percentage and buy volume for LPI activities across different blockchains.}
    \BeginAccSupp{ActualText={The figure shows a scatter plot showing the relationship between price increase percentage (x-axis) and buy volume (y-axis) for LPI activities across different blockchains.}}%
    \EndAccSupp{}
  \label{fig:price_pumping_effect}
\end{figure}
\mypara{High Price Impact with Minimal Investment.}
Fig.~\ref{fig:price_pumping_effect} shows the relationship between buy volume and price impact across the detected events revealing that manipulators achieved substantial price increases with minimal investment. Indeed, the median buy volume is merely $\$54$, with only $43.06\%$ (31) of events requiring more than $\$100$ in capital. Ethereum stands apart from other blockchains with its \$560 median, while others show median values below \$5.
Despite these modest volumes, the effects on the prices are dramatic: $50\%$ of manipulations generate price increases exceeding $500\%$, while $36.11\%$ produce gains surpassing $1,000\%$. In particular, the most extreme price distortions (exceeding $10,000\%$) occur with the smallest capital inputs, highlighting the vulnerability of low liquidity pools. The manipulation effectiveness varied significantly across blockchains: Base demonstrated the highest median price increase at $4 \times 10^{8}\%$, followed by Solana ($661\%$), BSC ($504\%$), and Ethereum ($214\%$).

\mypara{Few Actors Manipulation.}
The analysis of participant patterns further confirms the coordinated nature of these manipulations. 
On average, each LPI operation involves just 1.96 participants, with 66.67\% of events carried out by a single actor. This trend is consistent across all blockchains studied, highlighting how vulnerable low-liquidity pools are to price manipulation by individual actors.

\mypara{Targets Are Old Tokens.}
Unlike other manipulations that target newly launched tokens~\cite{cernera2023token,cernera2023ready}, LPI occurs predominantly at the more mature stages of the lifecycle of a token. Only 2 meme coins experienced manipulation within their first month, and only 9 within their first 90 days. The median time to manipulation is 209 days after token creation, suggesting that manipulators specifically seek out older tokens that never gained significant traction. These tokens present ideal targets as they appear more legitimate due to their longer existence while still maintaining the low liquidity pools necessary for effective price manipulation. This temporal pattern is particularly different on Solana, where tokens are targeted much earlier, with a median of just 115 days, compared to tokens on other blockchains, which have a median of over 163 days.

These results reveal that LPI is one of the most capital-efficient market manipulation strategies. With a relatively small investment, a single actor can drive significant artificial price increases, producing misleading price charts that attract unwary investors looking for high-growth opportunities.

\subsection{Legitimate vs. Manipulated Growth}
\label{subsec:artificial_growth_result}
The analysis of high-performing tokens shows that the vast majority display signs of market manipulation or conditions that facilitate it. By combining results from wash trading (287 tokens), LPI strategies (40), and strong anomaly indicators (412), we identify 586 tokens (82.89\% of all high-performing tokens) that exhibit at least one form of suspicious activity.
In particular, 120 tokens (20.48\%) show wash trading and strong anomaly indicators, with 95 tokens exhibiting extreme ownership concentration due to the top 10 accounts holding an average of 72.94\% of the supply.
Ownership concentration is even more pronounced in cases related to LPI, where of the 40 involved tokens, 32 show clear signs of concentrated ownership. This includes 27 tokens where the top 10 holders control an average of 80.62\% of the supply, and two cases where all the supply is acquired through bundle buy transactions, likely to obscure actual ownership.
Furthermore, we observe 12 tokens involved in both wash trading and LPI activities. This combination of tactics suggests coordinated efforts to inflate trading volume and price, creating a convincing illusion of strong market demand and growth.

This analysis leaves only 121 tokens that show no signs of market manipulation or suspicious ownership. In terms of blockchain distribution, these successful tokens are primarily on Solana (80 tokens), followed by BSC (22), Ethereum (12), and Base (7). When normalized by the total number of tokens on each platform for which we have price data (see Section~\ref{subsubsec:data_collection_results_performance}), Base has the highest success rate (with 1.14\%), followed by Solana (0.77\%), Ethereum (0.62\%), and BSC with the lowest success rate (0.30\%).



\section{Profit Extraction Manipulations}
\label{sec:profit_extraction}
With 82.8\% of high-performing meme coins showing manipulative practices (Sec.~\ref{subsec:artificial_growth_result}), this section explores whether such artificial growth is part of a calculated strategy to extract profits at the expense of investors.
To track this progression from manipulation to extraction, we extended our analysis using OHLCV data with hourly granularity from mid-January through mid-March 2025. 

\subsection{Strategic Exit via Rug Pulls}
\label{subsec:rug_pulls}
Rug pulls represent a severe form of exit scam where developers abandon their project after generating significant initial investment, typically by removing all liquidity from trading pools~\cite{zhou2024stop}. 
To detect rug pulls among high-performing tokens, we applied the labeling criteria and thresholds used in~\cite{mazorra2022not}:
(1) a single-day price drop of over 99\%, signaling that the token has effectively lost all value, a decline too extreme to be explained by normal market fluctuations, and
(2) a sustained drop in trading volume of more than 99\% in the following days, indicating that liquidity was removed and investors could no longer trade the token. These conditions confirm that the token has been abandoned and has not experienced a temporary market dip.
Applying this approach to our updated OHLCV data, we detect two tokens subjected to apparent rug pulls: one on Solana and the other on Ethereum. 
In particular, both tokens have previously undergone other market manipulations to generate artificial interest, one involved in wash trading and the other in LPI, indicating a progressive manipulation strategy culminating in the final exit scam.

\subsection{Orchestrated Pump and Dump Operations}
\label{subsec:pump_and_dump}
Pump and dumps are another common manipulation in which the price of a token is inflated through coordinated buying, followed by rapid selling that crashes the price and leaves late investors with substantial losses~\cite{xu2019anatomy}.
Our methodology for detecting pump and dump operations is the following:

\mypara{1. Price peaks detection:} We identify price peaks demonstrating overbought conditions with a Relative Strength Index exceeding 80, indicating excessive buying momentum~\cite{ctuaran2011relative}. For each detected peak, we examine the preceding 24-hour window to locate the local minimum price point as the potential manipulation starting point.

\mypara{2. Pump phase:} We classify a pump phase when three conditions are simultaneously met: (a) price increase exceeding 50\% from starting point to peak, (b) trading volume surge exceeding 500\% during this period, and (c) complete price ascent occurring within a maximum 24-hour timeframe. 

\mypara{3. Dump phase:} We then detect the subsequent dump phase through a sharp price decline exceeding 30\% following the peak, with post-dump average trading volume falling below 50\% of the volume observed during the pump phase, indicating rapidly waning interest after the orchestrated collapse.

We initially adopted the detection thresholds from~\cite{kamps2018moon}, who defined a pump as a 10\% price increase with a 400\% volume surge. However, as these parameters yielded excessive false positives, due to the volatility of meme coins, we implemented more conservative thresholds to prioritize precision over recall.
This approach detects 91 pump and dump operations affecting 60 different tokens, 54 on Solana, 3 on BSC, 2 on Base, and 1 on Ethereum. 
\begin{figure}
    \centerline{\includegraphics[width=0.49\textwidth]{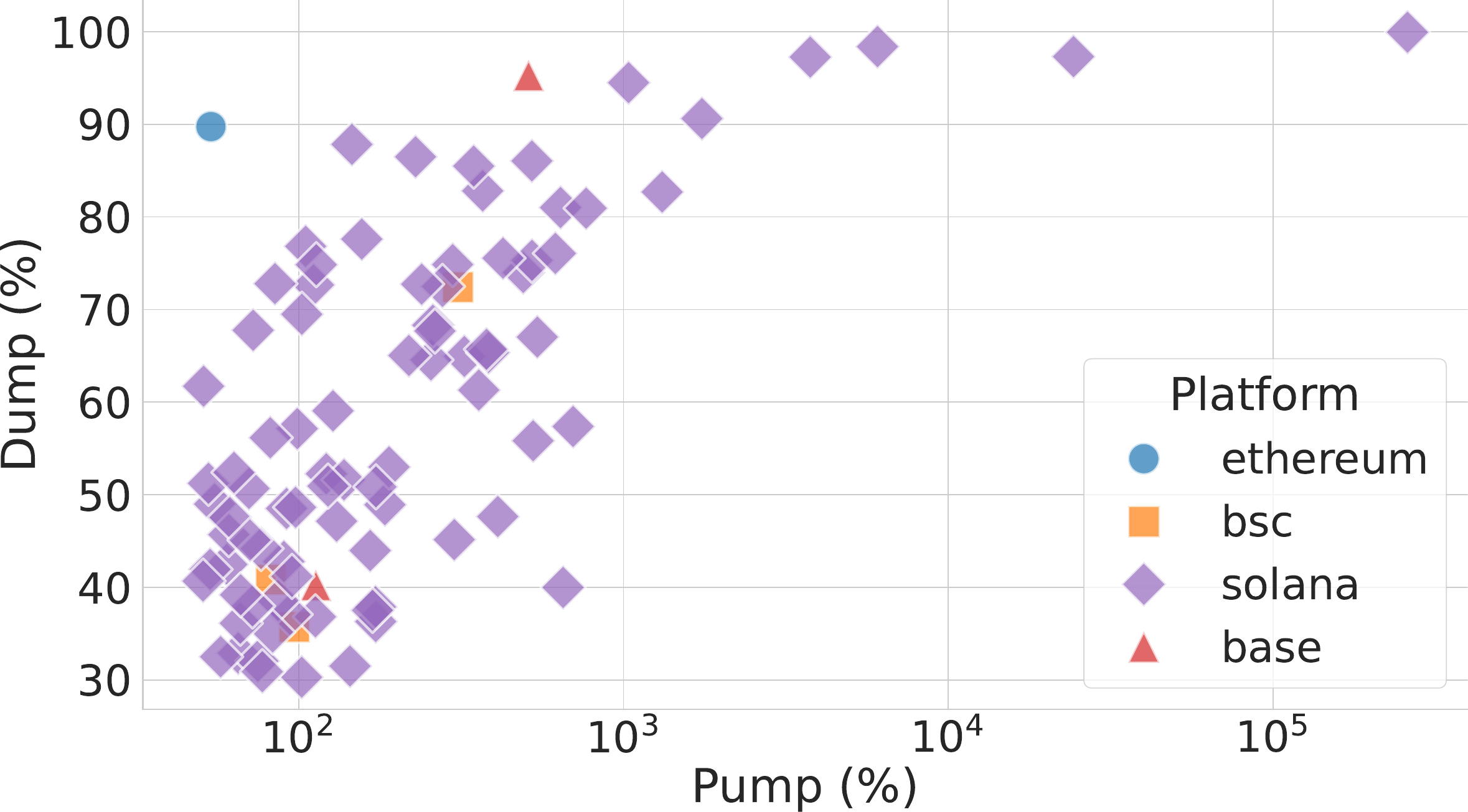}}
    \caption{
    Relationship between price increase during pump and decrease during dump across different blockchains.}
    \BeginAccSupp{ActualText={Scatterplot showing the magnitude of price increases during pump phases (x-axis, logarithmic scale) and subsequent price decreases during dump phases (y-axis) for each pump and dump operation, color-coded by blockchain.}}%
    \EndAccSupp{}
  \label{fig:pump_and_dump_phases}
\end{figure}
As shown in Fig.~\ref{fig:pump_and_dump_phases}, the pump phases exhibit significant variability in magnitude. The median price increase is 143.95\%, with 63.74\% of operations (58) featuring increases of at least 100\% and 7 operations (7.69\%) with increases exceeding 1,000\%.
The subsequent dump phases show correspondingly severe price collapses, with a median decrease of 55.82\%. In 59.34\% of cases (54), tokens lost at least 50\% of their value, while 17.58\% (16) experienced severe dumps exceeding 80\% price decline.

The 60 affected tokens reveal a concerning pattern: 61.67\% (37 tokens) were previously involved in artificial growth strategies, 35 in wash trading, and 3 LPI. If we also consider the rug pulled tokens, this ratio increases to 62.9\%.
This finding suggests that manipulations examined in Sec.~\ref{sec:growth_analysis} often function as preliminary steps in a broader scheme designed to attract investors before performing pump and dumps or rug pulls.

\subsection{Case Study: Delisted Tokens}
\label{subsec:delisted_tokens}
The removal (delisting) of tokens from major aggregators like CoinMarketCap and CoinGecko can signal either project abandonment or suspicious market behavior. To examine manipulations across this lifecycle, we analyze tokens that were once listed, having met the platforms' listing criteria, but were later removed.
These tokens represent an interesting case study, as they gained the credibility and visibility associated with major platform listings before their removal. We investigate whether these tokens used artificial growth to meet listing criteria and then exploited their increased visibility through market manipulation. We examine 124 meme coins delisted from CoinMarketCap and CoinGecko between October 2024 and March 2025 and report the results in Tab.~\ref{tab:delisted-meme-coins}. Among these, 84 tokens (67.74\%) showed signs of artificial growth, with 70 tokens exhibiting wash trading behavior, while 27 tokens employed LPI tactics. Additionally, 30 tokens (24.19\%) show evidence of profit extraction after listing: 23 were involved in pump and dumps and 7 in rug pulls. 
Of these, 26 tokens (86.67\%) had previously employed artificial growth tactics. Among the pump and dump tokens, 14 showed evidence of wash trading and 5 LPI. Likewise, all rug-pulled tokens showed signs of prior manipulation, with 5 involving wash trading and 6 LPI (4 of which exhibited both).

\begin{table}[h]
\centering
\small
\caption{Manipulation strategies in delisted meme coins}
\begin{tabular}{lcc}
\toprule
\textbf{Manipulation Category} & \textbf{\# Tokens} & \textbf{Percentage} \\
\midrule
Wash Trading & 70 & 56.45\% \\
LPI & 27 & 21.77\% \\
\textbf{Tokens with Artificial Growth} & \textbf{84} & \textbf{67.74\%} \\
\midrule
Pump and Dump & 23 & 18.55\% \\
Rug Pull & 7 & 5.65\% \\
\textbf{Tokens with Profit Extraction} & \textbf{30} & \textbf{24.19\%} \\
\bottomrule
\end{tabular}
\label{tab:delisted-meme-coins}
\end{table}

These results highlight a recurring manipulation strategy among delisted meme coins. Tokens first achieve legitimacy by leveraging artificial growth manipulations to meet listing thresholds. After gaining the visibility that comes with the listing, they execute exit strategies, revealing a coordinated effort to maximize gains before disappearing from the ecosystem.

\subsection{Economic Impact}
To quantify the financial harm posed by these manipulations, we analyzed the profit and loss flows for tokens involved in pump and dumps and rug pulls. Tab.~\ref{tab:economic_impact} summarizes the profits and losses across these events.

\begin{table}[t]
\centering
\small
\caption{Profit and Loss (PnL) for pump and dump (P\&P) participants and rug pull victims. 
\textit{Avg/Op} is the mean PnL per operation (in \$), \textit{Avg/Addr} is the mean PnL per wallet (in \$).}
\label{tab:economic_impact}
\begin{tabular}{lcccc}
\toprule
\textbf{} & \textbf{\#Addr.} & \textbf{Tot. PnL} & \textbf{Avg/Op} & \textbf{Avg/Addr}\\
\midrule
\textbf{P\&D} & & &\\
 Creators & 10,614 & 2.06M & 26,034 & 779\\
 Early Gainers & 3,245 & 3.93M & 51,652 & 579\\
 Late Gainers & 4,709 & 1.79M & 22,436 & 365\\
 \textit{Total Profit} & \textbf{18,568} & \textbf{7.78M} & \textbf{93,701} & \textbf{470}\\
 \hline
 Early Losers & 2,726 & -1.92M & -25,647 & -724\\
 Late Losers & 3,057 & -1.35M & -17,529 & -263\\
 \textit{Total Loss} & \textbf{5,783} & \textbf{-3.27M} & \textbf{-40,411} & \textbf{-617}\\
\midrule
\textbf{Rug Pull} & & &\\
 Victims & \textbf{11,368} & \textbf{-6.04M} & \textbf{-671,666} & \textbf{379}\\
\bottomrule
\end{tabular}
\end{table}

\mypara{Pump and Dump Impact.} 
We categorize pump and dump participants into three groups: creators, those who deployed the token or received it through airdrops, early investors, who purchased before the manipulation, and late investors, who bought during the pump phase.
Creators and early/late gainers extracted a combined \$7.78 million in profits. Conversely, investors who sold at a loss suffered \$3.27 million in realized losses. 
This figure is a conservative lower bound: It only accounts for realized losses, not considering the unrealized losses of victims still holding the now-devalued tokens.
The profit distribution reveals a clear exploitation hierarchy. Token creators extracted \$2.06 million across 10,614 addresses, averaging \$779 per wallet. This unusually high address count likely reflects the use of multiple wallets to obscure true ownership. Early investors who successfully exited accumulated \$3.93 million across 3,245 addresses (\$579 per wallet), while late investors achieved \$1.79 million across 4,709 addresses (\$365 per wallet). In total, 18,568 addresses realized gains averaging \$470 per wallet.
However, these gains came at the expense of 5,783 addresses that suffered realized losses averaging \$617 per wallet. Early investors who sold at a loss experienced \$1.92 million in total losses (\$724 per wallet), while late investors suffered \$1.35 million (\$263 per wallet). 

\mypara{Rug Pulls Impact.}
Unlike pump and dumps where we track only realized losses, rug pulls effectively trap all remaining holders, as liquidity is removed from the market. Thus, we identify victims as addresses that purchased tokens before the rug pull and were unable to sell.
The total losses from rug pulls reached \$6.04 million, with an average of \$671,666 per event, and with the most destructive operation causing losses exceeding \$2.89 million. In particular, rug pulls affected a total of 11,368 addresses, with an average of 1,263 addresses per event and one operation alone impacting 4,380 wallets.

Overall, we document a total financial loss exceeding \$9.3 million, impacting over 17,000 victim addresses. This highlights that the harm from these manipulations is widely distributed among retail investors, while the profits remain concentrated in the hands of creators and insiders, who likely employ multiple addresses to conceal their ownership concentration (as shown in Sec.~\ref{subsubsec:ownership_concentration}).

\section{Discussion}
\mypara{Implications for Ecosystem Stakeholders.}
Our findings reveal critical deficiencies in how cryptocurrency platforms present market data and the inadequacy of existing protection mechanisms. Security metrics, such as token concentration, require specialized knowledge to interpret and are typically buried at the bottom of token pages. Most critically, platforms fail to alert users to artificial growth strategies, relying instead on volume and liquidity metrics that our results demonstrate can be easily manipulated. To maintain their role as reliable information sources, these platforms must implement two key reforms. First, risk indicators should be displayed prominently using accessible language rather than technical metrics. Second, established aggregators must expand verification beyond basic technical checks to include continuous monitoring for artificial growth, which our analysis shows frequently precedes exploitation schemes.
Moreover, this work supports regulators enforcing frameworks like MiCA~\cite{mica} by providing a comprehensive analysis of the manipulations, identifying detection patterns, quantifying the economic prevalence, and revealing novel strategies, such as LPI.

\mypara{Countermeasures.}
Exit schemes such as rug pulls and pump-and-dumps execute rapidly, leaving minimal reaction time. Nonetheless, our work reveals that 62.9\% of tokens subjected to profit extraction (86.67\% among delisted tokens) had previously undergone artificial growth. This temporal pattern enables a viable defensive strategy: Detecting wash trading and LPI activities establishes an actionable warning window before profit extraction occurs. In this study, we employed conservative thresholds to minimize false positives, meaning our results likely underestimate total manipulation volume. Operational systems must carefully calibrate the precision-recall trade-off: Aggressive detection may identify more manipulations but risks flagging legitimate tokens, thereby undermining user trust, while overly conservative detection may fail to protect investors. 
Beyond technical detection, the most robust countermeasure remains investor education. Informed users who can recognize manipulation mechanics are unlikely to provide the exit liquidity that enables these schemes.

\section{Conclusion and Future Work}
Meme coins have rapidly expanded into the cryptocurrency market. However, their community-driven nature makes them highly susceptible to manipulation. This study presents the first cross-chain analysis of meme coins across four major blockchains, examining how different manipulation strategies are combined and executed sequentially rather than studying them in isolation.
We found that 82.8\% of high-performing meme coins exhibited clear signs of artificial growth, including wash trading and LPI, and concentrated token ownership. This result suggests that the dramatic gains often associated with meme coins are more likely the result of coordinated manipulation than organic market dynamics.
We also detected pump and dumps and rug pulls, with the majority of the tokens involved having previously experienced artificial growth, indicating that early manipulations often serve as precursors to larger exploitation strategies. Finally, we documented this pattern also in tokens delisted from established aggregators, discovering that 67.74\% showed signs of artificial growth before delisting, and 24.19\% exhibited profit extraction tactics after gaining visibility through listing. 
Crucially, we quantified the impact of these exit schemes, linking them to over 17,000 victim addresses and cumulative realized losses exceeding \$9.3 million.
These results challenge the perception of meme coins as harmless speculation, revealing instead a structured environment enabling systematic wealth transfer from uninformed to manipulative actors.

This work underscores the urgent need for stronger investor protection measures in the meme coin ecosystem. 
Future work should focus on developing real-time detection systems capable of identifying wash trading and LPI operations, as well as tracking how these evolve into profit extraction schemes. 
Finally, to gain a deeper understanding of these strategies, it would be valuable to examine the off-chain promotional activities of projects involved in market manipulation. This includes analyzing their websites and social media to study how online engagement is coordinated with the timing of manipulative activities.

\section*{Acknowledgments}
This work has been partially funded by MUR National Recovery and Resilience Plan, SERICS (PE00000014).



\cleardoublepage
\appendix

\section*{Ethical Considerations}

\subsubsection*{Data Collection and Privacy.}
This study relies exclusively on publicly available data collected from open and accessible sources, including blockchain explorers (\eg Etherscan, BscScan, Solana RPC), cryptocurrency tracking platforms (\eg CoinGecko, CoinMarketCap), decentralized exchange aggregators (\eg DexScreener), and blockchain analytics services (\eg Dune Analytics). All data pertain to smart contracts, token transactions, and aggregated market information; the study did not involve any interaction with human subjects or the collection of personally identifiable information (PII). Furthermore, throughout our investigation, we made no attempts to correlate blockchain addresses with external off-chain information that could compromise user privacy.
In particular, data collection was conducted responsibly. All scraping scripts were designed to respect the operational stability of the target platforms, with requests spaced by at least five seconds to avoid overloading any external service. Additionally, no financial transactions were made during this study, and the authors did not engage in the trading, creation, or promotion of any meme coins examined in this work. 

\subsubsection*{Positive Impacts.}
Several stakeholders benefited by this research. 

\mypara{Investors \& End-Users.}
The primary beneficiaries are current and potential investors. Our work provides two key benefits: education and direct protection. It serves as a large-scale, empirical warning, moving beyond anecdotal evidence to show that manipulation is widespread and systemic, especially in high-performing tokens. Additionally, our detection methods are designed to be lightweight and suitable for integration by third-party platforms to provide real-time risk scores, protecting users from investing in tokens that are actively being manipulated.

\mypara{Information Aggregators.}
Our findings also provide information aggregators with new, data-driven heuristics to improve their systems. For \textit{crypto data aggregators} like CoinMarketCap and CoinGecko, which require tokens to meet specific criteria, our research reveals how malicious actors exploit this trust using artificial growth strategies like wash trading and Liquidity Pool-Based Price Inflation (LPI). Our work provides these platforms with detection tools to check tokens more rigorously before listing, thereby maintaining the trust upon which their business models rely. Concerning \textit{DEX aggregators} like DexScreener and CoinSniper, these platforms list new tokens almost instantly, making them essential for "discovering" new assets. Their problem is twofold: First, their security metrics (\eg token ownership concentration) are often buried at the bottom of the page and are difficult for non-expert users to understand; second, their warnings are often generic and insufficient (\eg "this token could be a scam... DYOR"). 
By adopting our specific detection methods, these platforms could display prominent, actionable warning banners, such as flagging signs of LPI, at the top of token pages. This would allow them to differentiate themselves as safer platforms for discovering new assets.

\mypara{The S\&P Community.} 
We make three key contributions for the community: (1) We present a novel, large-scale dataset focused on meme coins across multiple blockchains, providing a foundation for future studies in this area; (2) We define and provide a detection method for a new manipulation technique (LPI); (3) We highlight the need for future work to study manipulations in combination rather than as isolated events. 

\mypara{Regulators.}
As regulators begin to deploy new frameworks like MiCA to combat market manipulation (Art. 91~\cite{MiCA2023_Art91}), our work shows how prevalent and how systemic these manipulations are, identifying also novel manipulation techniques (LPI). Our findings are therefore essential for regulators to understand the specific, on-chain mechanisms they will need to detect in order to successfully enforce anti-abuse provisions.

\subsubsection*{Negative Impacts}
\mypara{Malicious Actors (Manipulators).}
We acknowledge the dual-use nature of this research. By detailing the mechanisms of novel or complex manipulations like LPI, there is a risk of educating new or less-sophisticated actors about techniques they may not have been aware of. Furthermore, for already-sophisticated manipulators, our paper could be seen as a guide to refining their attacks to evade the specific detection heuristics we propose.

\subsubsection*{Mitigation and Decision to Publish}
The manipulation techniques we describe, particularly LPI, are already in widespread use, as our study demonstrates: Over 17,000 addresses suffered quantifiable harm, including \$3.27M in realized losses from pump-and-dump schemes and \$6.04M in direct losses from rug pulls.
This knowledge asymmetry currently favors malicious actors who operate with impunity, while victims and platforms lack detection tools. Public disclosure shifts this balance. 
Our paper serves to arm the defenders, users, platforms, regulators, and researchers, with the same knowledge, which is the first and most critical step toward building a defense. 

By publishing this work, we enable immediate platform protection, evidence-based regulatory action, advanced detection research, and investor education. Thus, we conclude that the clear harm to investors and the defensive benefits of disclosure substantially outweigh the marginal risk of informing actors who largely already possess this knowledge.


\section*{Open Science}
In adherence to open science principles and to support transparency and reproducibility of our findings, we publicly release on Zenodo (\url{https://doi.org/10.5281/zenodo.17830943}) all datasets and results of this study, as well as the methodological implementations employed to detect anomalies, artificial growth operations (wash trading and LPI activities), and profit extraction schemes (rug pulls and pump and dumps).

Specifically, the repository includes the complete dataset of 34,988 confirmed meme coins (Sec.\ref{subsubsec:dataset_refinement}), the set of tokens validated across Ethereum, BNB Smart Chain, Solana, and Base (Sec.~\ref{sec:address_validation}), and data about creation date of the smart contract (Sec.~\ref{sec:temporal_evolution_of_memecoins}). It also contains token price data spanning the three-month observation window, which support the price performance analysis presented in Sec.~\ref{sec:performance_analysis}.

In addition, we provide labeled subsets of tokens used throughout the analyses: high-performing meme coins, delisted tokens, and those exhibiting signs of suspicious behavior, including ownership concentration anomalies, artificial growth operations (\eg wash trading and Liquidity Pool-Based Price Inflation, or LPI), and profit extraction schemes such as rug pulls and pump-and-dump events (Sec.~\ref{sec:growth_analysis} and Sec.~\ref{sec:profit_extraction}).

The dataset is complemented by OHLCV (Open, High, Low, Close, Volume) data at both daily and hourly granularity, which were used to detect manipulations.
We also release the transaction-level data and SQL queries used to extract it from Dune Analytics, which we leveraged to validate suspicious activities related to wash trading and LPI (Sec.~\ref{subsec:advanced_detection_wash_trading} and Sec.~\ref{sec:lpi_detection}). 

Finally, we release a complete set of Python notebooks implementing the core manipulation detection methods described in the paper, from anomaly identification to the sequential analysis of artificial growth manipulations and profit extraction strategies.

\newpage
\section*{Appendix}

\begin{figure}[!htbp]
    \centerline{\includegraphics[width=0.49\textwidth]{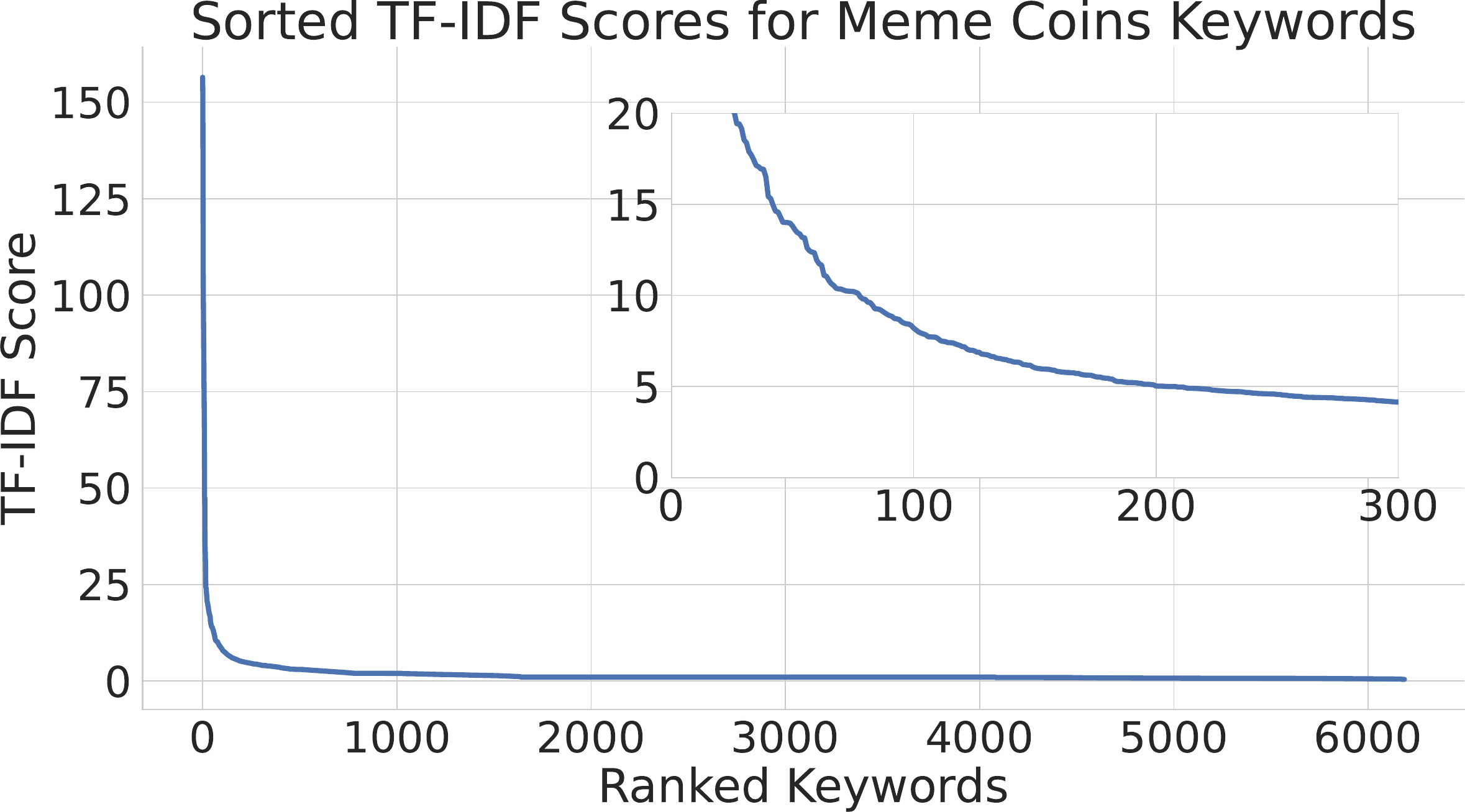}}
    \caption{TF-IDF scores of keywords associated with meme coin names.}
    \BeginAccSupp{ActualText={A plot showing the TF-IDF scores of keywords extracted from meme coins' names.}}%
    \EndAccSupp{}
  \label{fig:tf_idf_names}
\end{figure}

\begin{figure}[!htbp]
    \centerline{\includegraphics[width=0.49\textwidth]{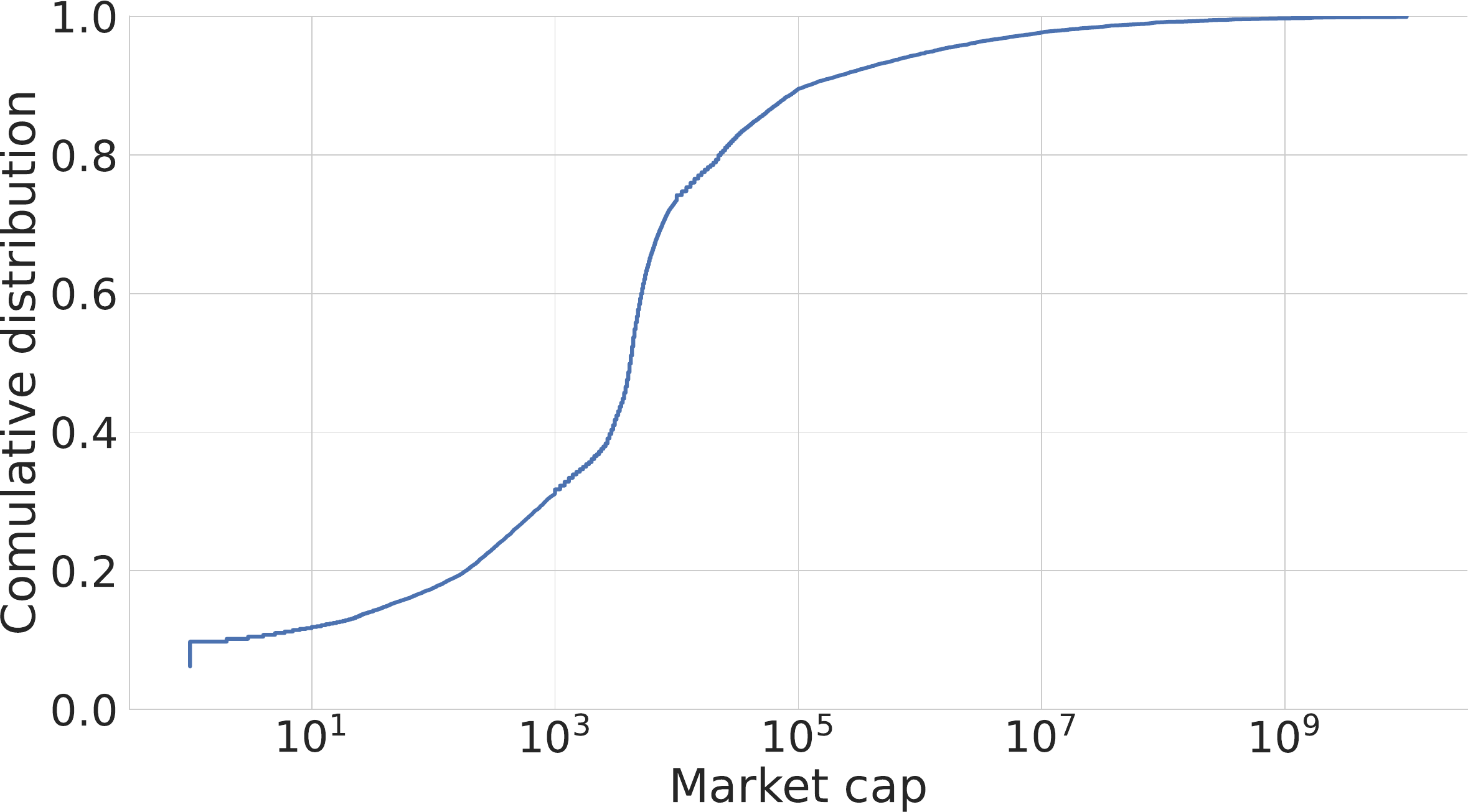}}
    \caption{CDF of the market capitalization of meme coins collected.}
    \BeginAccSupp{ActualText={A plot showing the CDF of the market capitalization of meme coins collected.}}%
    \EndAccSupp{}
  \label{fig:cdf_dataset_refinement}
\end{figure}

\cleardoublepage
\bibliographystyle{plain}
\bibliography{main}

\end{document}